\documentclass[fleqn,usenatbib]{mnras}

\usepackage{newtxtext}
\usepackage{newtxmath}

\usepackage[T1]{fontenc}
\usepackage{ae,aecompl}

\usepackage{graphicx}	
\usepackage{amsmath}	
\usepackage{amssymb}	
\usepackage{xcolor}     
\usepackage{float}      
\usepackage[normalem]{ulem}

\newcommand{\eg}{\textit{e.g. }}
\newcommand{\ie}{{\it i.e.,}\ }

\title[Detecting episodes of star formation 
using Bayesian model selection]{Detecting episodes of star formation 
using Bayesian model selection}

\author[A. J. Lawler et al.]{
Andrew J. Lawler,$^{1}$\thanks{E-mail: lawler\_andy\_j@gmail.com}
Viviana Acquaviva$^{2,3}$
\\
$^{1}$Department of Statistics, Baylor University, One Bear Place \#97140, Waco, TX 76798, USA\\
$^{2}$Department of Physics, CUNY NYC College of Technology, 300 Jay Street, Brooklyn NY 11201, USA\\
$^{3}$Institut de Ci\`encies del Cosmos, Universitat de Barcelona, Carrer Mart\`i i Franques 1, 08028 Barcelona (Spain).\\
}

\date{Accepted XXX. Received YYY; in original form ZZZ}

\pubyear{2020}

\begin{document}
\label{firstpage}
\pagerange{\pageref{firstpage}--\pageref{lastpage}}
\maketitle

\begin{abstract}
Bayesian model comparison frameworks can be used when fitting models to data in order to infer the appropriate model complexity in a data-driven manner. We aim to use them to detect the correct number of major episodes of star formation from the analysis of the spectral energy distributions (SEDs) of galaxies, modeled after 3D-HST galaxies at $z \sim$ 1. Starting from the published stellar population properties of these galaxies, we use kernel density estimates to build multivariate input parameter distributions to obtain realistic simulations. We create simulated sets of spectra of varying degrees of complexity (identified by the number of parameters), and derive SED fitting results and evidences for pairs of nested models, including the correct model as well as more simplistic ones, using the \textsc{bagpipes} codebase with nested sampling algorithm \textsc{MultiNest}. We then ask the question: is it true - as expected in Bayesian model comparison frameworks - that the correct model has larger evidence?

Our results indicate that the ratio of evidences (the Bayes factor) is able to identify the correct underlying model in the vast majority of cases. The quality of the results improves primarily as a function of the total S/N in the SED. We also compare the Bayes factors obtained using the evidence to those obtained via the Savage-Dickey Density Ratio (SDDR), an analytic approximation which can be calculated using samples from regular Markov Chain Monte Carlo methods. We show that the SDDR ratio can satisfactorily replace a full evidence calculation provided that the sampling density is sufficient.
\end{abstract}




\section{Introduction}
\label{sec:intro}
The spectral energy distribution (SED) of a galaxy encodes a vast amount of information about the galaxy's physical properties. Complex processes such as the formation and evolution of stars and the chemical composition of the interstellar medium (ISM) between the stars dominate the emission from different parts of the electromagnetic spectrum. These processes are often highly interdependent, making it difficult to isolate the contribution of each component separately (\eg \citealt{McKee2007, Walcher2011, Conroy2013}). 

Past and ongoing photometric and spectroscopic surveys, such as SDSS (\citealt{York2000}), COSMOS (\citealt{Scoville2007}), 2MASS (\citealt{Skrutskie2006}), UltraVISTA (\citealt{McCracken2012}), CANDELS (\citealt{Grogin2011}), and 3D-HST (\citealt{Skelton2014}) provide measurements from the ultraviolet to infrared wavelengths in a range of redshifts for millions of galaxies.  To interpret these data, a variety of models to describe these physical processes and their interplay have been developed in the past two decades. In parallel with the improvement in data quality and volume, models have grown increasingly sophisticated.

Among the many components of these models there are, for example, the initial stellar mass function (IMF), the star formation history (SFH), the dust attenuation law, the chemical enrichment history, and assumptions about the underlying stellar population models. In particular, the SFH of a galaxy has received a great deal of attention because it provides important insights on which mechanisms dominate galaxy formation and evolution. 

The method of SED fitting compares libraries of models to spectro-photometric data to derive physical properties of interest. Techniques such as $\chi^2$ fitting have historically been used to identify a best-fit model and estimate parameters (\citealt{Arnouts1999, Bolzonella2000}). More recently, Bayesian methods employing Markov chain Monte Carlo (MCMC) sampling algorithms have been employed to better constrain the correlated parameter spaces of galaxy models and analyze posterior probability distributions for parameter inference (\eg \citealt{GalMC}).

Bayesian model comparison is a statistical technique aimed at determining the optimal model complexity that is warranted by the data. Despite being used widely in other fields of astronomy, for example in the context of cosmological model selection (\citealt{Mukherjee2006, Shaw2007, Trotta2007, Trotta2008}), it has seldom been applied to SED fitting thus far. \citealt{Han2012} used \textsc{MultiNest} to calculate and compare the Bayesian evidence from the SED models for hyperluminous infrared galaxies; the same authors used the evidence to  compare two popular stellar  population synthesis (SPS) models in \citealt{Han2014, Han2018}. \citealt{Dries2016, Dries2018} used Bayesian evidence in a hierarchical setting to compare the prior spaces for several choices of IMFs. \citealt{Salmon2016} used Bayesian evidence to test dust laws on a sample of galaxies in CANDELS. 

One of the crucial issues in SED fitting is that galaxies are extremely complicated objects, and modeling them with a few parameters is a necessary but very crude simplification. Until recently, galaxies have typically been described as combinations of simple stellar population models of fixed metallicity and stellar ages. The process of stellar assembly was often described using the exponentially declining ($\tau$) model, which assumes that galaxies form stars at a declining rate, parameterized by the half-time $\tau$, and the attenuation due to dust was often described by a one parameter ``screen" model (\eg \citealt{Calzetti2000}). However, more recently, it has been become clear that these extremely simplified assumptions are unable to capture the complexity of galaxies' behaviors, and also cause biases in the determination of the physical properties of galaxies. In particular, it has been shown that wrongly reconstructed star formation histories introduce noticeable biases in many parameters that are usually estimated through spectral energy distribution fitting, such as stellar masses, stellar age indicators, dust content, and redshift (\eg \citealt{Mobasher2015, Pacifici2015, Iyer2017, Leja2017}). In \citealt{Acquaviva2015}, we evaluated the impact of different sources of non-algorithmic systematics on the recovered SED fitting parameters and concluded that a wrong star formation history is the most harmful. Similarly, \citealt{Iyer2017} and \citealt{Carnall2018} found that fitting the SFH using single stellar populations and simple functional forms (\eg exponentially declining or constant models) leads to a bias of up to $70\%$ in the recovered physical properties of the galaxies.

Progress in the modeling of stellar populations has been more promising, with several groups developing sophisticated algorithms used to generate and fit simulated spectra in the ultraviolet-through-infrared range with increased flexibility (\citealt{daCunha2008, Conroy2009, Chevallard2016, Carnall2018, Leja2017, Iyer2019}). This means that we are now technically able to explore much larger parameter spaces, but are subject to the curse of dimensionality: the process of SED fitting might take a much longer time, which does not suit large upcoming surveys, and/or the data might not be high enough quality to resolve the many degeneracies inherent to models with a large number of parameters.

For these reasons, in this paper we set out to determine whether Bayesian model selection can be used to infer \textit{from the data} what is the true model complexity. Because the choice of star formation history is crucial to this issue, our first exploratory question is: are we able to successfully detect the correct number of major episodes of star formation, which is an index of model complexity, by comparing the evidence values of the different models? We will use simulated spectro-photometric data to answer this question under different assumptions, and then validate the method.

While in principle the Bayesian model comparison method is applicable to models that can occupy different prior spaces, in practice the relevant calculations depend on both the volume occupied by the likelihood and the volume occupied by priors. As a result, it is difficult to disentangle the contribution of the data and the priors to the model comparison. For this reason, we use \textit{nested} models: a hierarchical setting where we compare a more complicated $N + k$ parameter model to a simpler N parameter model. Comparison between these two models is achieved by the use of the Bayes factor, which is simply the ratio of the Bayesian evidences for each respective model. The mechanics of the Bayes factor can be interpreted as the factor by which the model space shrinks when the data arrives and will only favor a model with more parameters when the data warrants the additional complexity (\eg  \citealt{mackay}). 

This paper is organized as follows. In Sec. \ref{sec:data} we describe the data, inferred parameters, and noise profile used to construct our mock data. This includes a novel use of kernel density estimation to derive realistic multivariate distributions for the parameter space, as well as the error structure. We then review the SED fitting code \textsc{bagpipes}, which is used to generate a realistic galaxy catalog and to fit the models. We make a number of assumptions about the SFH, IMF, and dust law used for the galaxy population and define the prior space used. In Sec. \ref{sec:evidence} we review Bayesian inference methods and summarize \textsc{MultiNest}, the sampling algorithm used for both parameter estimation and evidence calculation. In Sec. \ref{sec:bayesfactors} we define the SDDR ratio, discuss its expected range of applicability, and compare the two approaches used to calculate the Bayes factor. In Sec. \ref{sec:results} we describe the simulated scenarios that we explore, which includes analyzing combinations of various design points to assess the effectiveness of the Bayes factor as a model comparison tool, and discuss our results. Our conclusions are summarized in Sec. \ref{sec:conclusions}.

\section{Data}
\label{sec:data} 
We apply our methodology to simulated galaxies modeled after those in the CANDELS GOODS-South field from v4.1 of the 3D-HST catalog; this ensures that we employ realistic distributions for input parameters and observational errors. The main catalog consists of ultraviolet to mid-infrared wavelength measurements of 50,507 galaxies, and it includes fitted stellar population parameters obtained through FAST (\citealt{Kriek2009}).

We select objects according to the following criteria:

\begin{enumerate}
\item Objects with the \texttt{use\char`_phot} flag set to 1. This is defined as an object not classified as a star or close to a bright star, is well-exposed on the F125W and F160W bands, has a S/N $>$ 3 in F160W, and has both a reasonable photometric redshift fit and a reasonable stellar population fit according to the EAZY/FAST codes. 
\item Objects with the \texttt{z\char`_peak} flag between 0.9 and 1.1, which is defined as the peak of the photometric redshift distribution according to the EAZY code. 
\end{enumerate}

The above criteria results in a reference catalog of 4,567 galaxies; we use a total of 23 bands for each galaxy. The parameter values for this sample estimated by FAST and reported by the 3D-HST collaboration are used as the basis for the kernel density estimation of the input parameter space. We also use kernel density estimation methods for creation of the mock catalog error structure, as described in \ref{sec:KDE} below.

\subsection{Kernel Density Estimation}
\label{sec:KDE}

In order to generate a realistic parameter space for SED fitting, we choose to leverage kernel density estimates (KDEs) of the fitted stellar population parameters from FAST, which include stellar mass, stellar age, dust extinction, photometric redshift, and the timescale of star formation ($\tau$), as the FAST code assumes an exponentially declining model. Since these parameters were fitted on a grid of possible values, there are gaps in the fits that limit the precision of the estimated values. However, our multivariate kernel density estimate approach allows us to build a realistic population of parameter values on a continuum while correctly capturing the correlations among the parameters.  

Kernel density estimation is a non-parametric data smoothing technique that uses a bandwidth parameter to approximate the probability distribution function of a data sample. We use the k-nearest neighbor (kNN) density estimator in \texttt{scikit-learn} (\citealt{Pedregosa2011}), where for a given point $x$, the multivariate kNN density estimator estimates the density by (\citealt{Tran2009})

\begin{equation} \label{eq:KDE1}
\hat{f}_{\text{knn}}(\textbf{x}) = \frac{1}{n V_k(\textbf{x})} \sum_{i=1}^n K \left( \frac{\textbf{x} - X_i}{\textbf{H}} \right)
\end{equation}

\noindent where $V_k(\textbf{x}) = c_d r_k^d(\textbf{x})$ is the volume of the $d$-dimensional sphere, $r_k^d(\textbf{x})$ is the Euclidean distance from $\textbf{x}$ to the $k$-th nearest neighbor (in other words, the radius of the sphere), and $c_d$ is the constant factor applied (i.e. $c_1 = 2, c_2 = \pi, c_3 = \frac{4 \pi}{3}$, etc.). Also, $K(\cdot)$ is a multivariate kernel, and $\textbf{H}$ is a vector of bandwidths $h$, which are all equal in this case. The generalized Euclidean distance metric is given by

\begin{equation} \label{eq:KDE2}
d_{i,j} = \sqrt{\sum_{k=1}^p (x_{ik} - x_{jk})^2 }
\end{equation}

We use a multivariate standard normal kernel, given by 

\begin{equation} \label{eq:KDE3}
K \left( \frac{\textbf{x} - X_i}{\textbf{H}} \right) = (2 \pi)^{-(d/2)} \text{exp} \left(- \frac{(\textbf{x} - X_i)^T(\textbf{x} - X_i)}{\textbf{2H}} \right)
\end{equation}

This kernel is chosen as it has the desirable properties of being symmetric and unimodal. Several of the parameter distributions reported by FAST are also approximately log-normal, making this kernel suitable for density estimation.

\begin{figure}
\centering
	\includegraphics[width=\columnwidth]{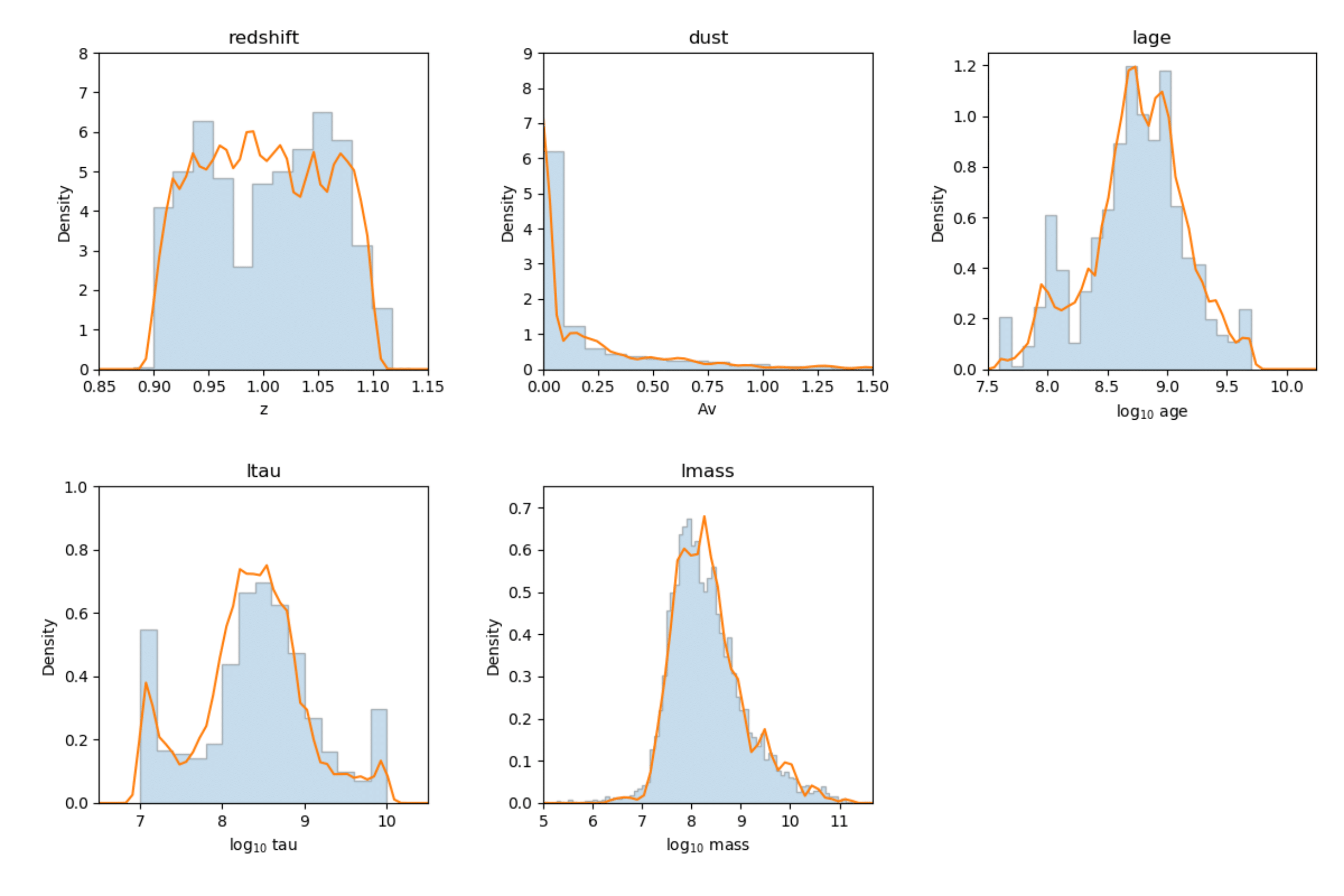}
    \caption{The distributions of the FAST parameter values from 4,567 galaxies in the GOODS-S field of the 3D-HST catalog are represented by the blue histograms.  The marginalized KDE of each parameter distribution from the multivariate KDE are represented by the orange lines. A bandwidth of 0.0861 is used for the multivariate KDE, therefore the marginalized KDEs also have a bandwidth of 0.0861. Note that both the histograms and the marginalized KDEs are normalized.}
    \label{fig:FAST_Params_KDE_Hist_Panel_3_2_z1}
\end{figure}

The bandwidth plays a crucial role in determining the bias-variance relationship of the resulting distribution. A larger bandwidth results in a smoother distribution with less variance and more bias, whereas a smaller bandwidth results in an less smooth distribution with more variance and less bias. The bandwidth parameter is chosen using a grid-based search with 5-fold cross-validation. Outliers above the 99.5 percentile in the right tail of each FAST parameter distribution are removed so as not to overestimate the tails of the kernel density estimates. The total log-likelihood of the data in each test set is calculated for each bandwidth, and the bandwidth is chosen according to the largest mean log-likelihood. Marginalized distributions for each parameter in the multivariate KDE are shown in Fig. \ref{fig:FAST_Params_KDE_Hist_Panel_3_2_z1}. The agreement between the histograms of estimated parameters and the probability distributions derived from the KDE is very good; furthermore, this method preserves all the correlations between different parameters, ensuring that the input parameter space of our simulated galaxies is fully realistic. 

\begin{figure}
\centering
	\includegraphics[width=\columnwidth]{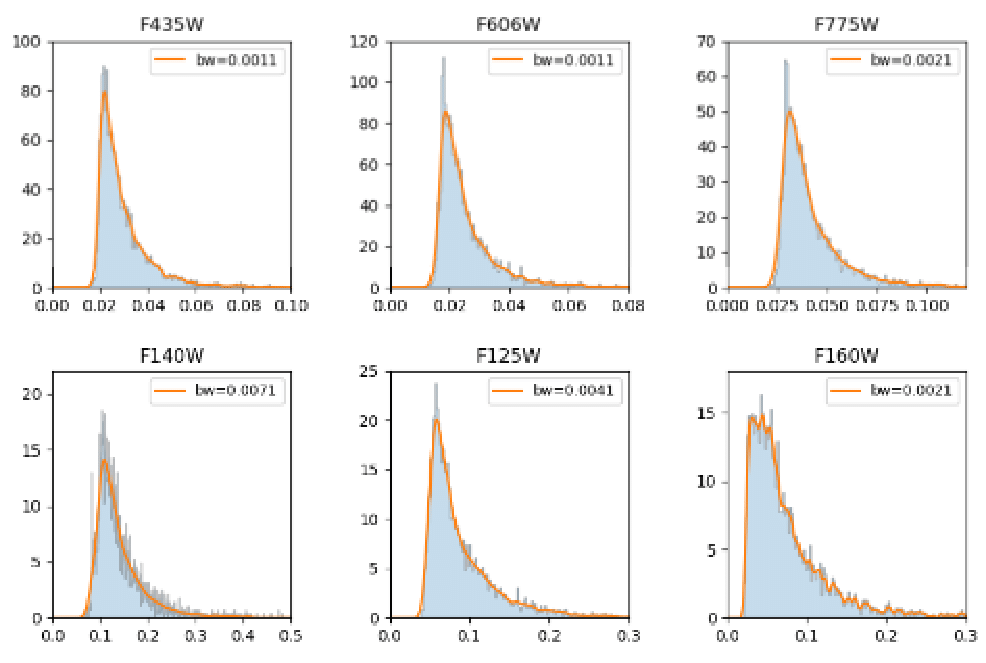}
    \caption{A selection of the distributions of the observational errors from 4,567 galaxies in the GOODS-S field of the 3D-HST catalog are represented by the blue histograms. The marginal KDE of each distribution from the multivariate KDE are represented by the orange lines. Each KDE has a bandwidth chosen by 5-fold cross validation.}
    \label{fig:Obs_Error_KDE_Hist_Panel_5_5_z1_subset}
\end{figure}

Similarly, we create KDEs of the distributions of observational errors. However, these are not treated in a multivariate way since errors are assumed to be independent across observations. As with the multivariate parameter KDE, values above the 99.5 percentile are removed before each error KDE is constructed. These are shown in Fig. \ref{fig:Obs_Error_KDE_Hist_Panel_5_5_z1_subset}.

\subsection{Generating Mock Data}
\label{sec:gen_mock_data}

Armed with our realistic multivariate distribution of galaxy properties, we are now ready to generate our mock catalogs. We use version 0.7.9 of \textsc{bagpipes} (\citealt{Carnall2018}), a modern SED fitting code that we chose for its sophistication in  modeling galaxies and its ability to infer parameters using nested sampling, which will be crucial for the evidence calculations in model comparison. 

In recent years more complex parametric models, such as the delayed exponentially declining tau model with additional stochastic bursts of star formation, have emerged as a more realistic alternative to the exponentially declining tau model (see \eg \citealt{Leja2019} and references therein).

For our main mode of star formation, we use the delayed exponentially declining tau model, defined as (\citealt{Gavazzi2002}; \citealt{Behroozi2010}; \citealt{Lee2010})

\begin{equation} \label{eq:delayed_tau}
\text{SFR}_{\text{delayed}}\,(t, t_0, \tau) \propto (t-t_0) \,  e^{-\frac{t-t_0}{\tau}},
\end{equation}

\noindent where $t$ is the age of the universe at observation, $t_0$ is the time of the onset of star formation, and $\tau$ is the width of the star formation history. 

Random samples from the multivariate KDE based on the FAST parameter distributions (described in Sec. \ref{sec:KDE}) are used as parameter inputs for a delayed tau model.

When present, a second stellar population (burst) is modeled as a simple delta function in time, with parameters burst age and mass formed in the burst. In this case, we uniformly sample a burst mass of 10\%-90\% of the total mass (effectively splitting the randomly drawn stellar mass sample from the multivariate KDE into two components) and we uniformly sample the random burst age between 0.05 and 0.90 Gyr.

We assume a Calzetti \citep{Calzetti2000} dust law, parameterized by the attenuation in the V band ($A_v$), with an additional multiplicative factor $\eta$ for stars in birth clouds of age $<$ 0.01 Gyr. We set the ionization parameter $\text{log}_{10} \ U$ for the nebular component to $-3.0$, and we assume a Kroupa \& Boily (2002) initial mass function (IMF). 

A summary of all the \textsc{bagpipes} parameter inputs is provided in Table \ref{tab:bagpipe_params}.

\begin{table}
\begin{center}
\begin{tabular}{l r r}
  \hline
  \textsc{bagpipes} Parameter & Value & Range\\
  \hline
  Global Components & \\
  redshift & KDE & 0.9-1.1\\
  t\_bc & 0.01 & \\
  \hline
  Dust Components & \\
  law & Calzetti & \\
  $A_{\text{v}}$ & KDE & 0-3.0\\
  $\eta$ & 2.0 & \\
  \hline
  Nebular Components & \\
  logU & -3.0 & \\
  \hline
  SFH Components & \\
  log$_{10}$(mass$_{\text{delayed} \ \tau}$) & KDE*random & 6.5-12.5 M/M$_\odot$\\
  log$_{10}$(mass$_{\text{burst}}$) & KDE*random & 6.5-12.5 M/M$_\odot$\\
  metallicity & 0.02 & \\
  log$_{10}$(age$_{\text{delayed} \ \tau}$) & KDE & 7.7-9.7 Gyr\\
  age$_{\text{burst}}$ & random & 0.10-0.90 Gyr\\
  $\text{log}_{10}(\tau)$ & KDE & 7.0-10.0 Gyr\\
  \hline
\end{tabular}
\caption{\protect\textsc{bagpipes} parameter inputs used to generate mock data catalog. Parameters denoted by ``KDE" are drawn from a multivariate input distribution obtained by using kernel density estimation. For stellar mass, ``KDE*random" indicates that the total mass is obtained through the KDE method, and the mass of the secondary population is obtained as a uniform random number in the (0.1, 0.9) interval.}
\label{tab:bagpipe_params}
\end{center}
\end{table}

\subsection{Mock Data vs 3D-HST Data}
\label{sec:mock_data_3dhst_data}

\begin{figure*}
\centering
	\includegraphics[width=0.8\textwidth]{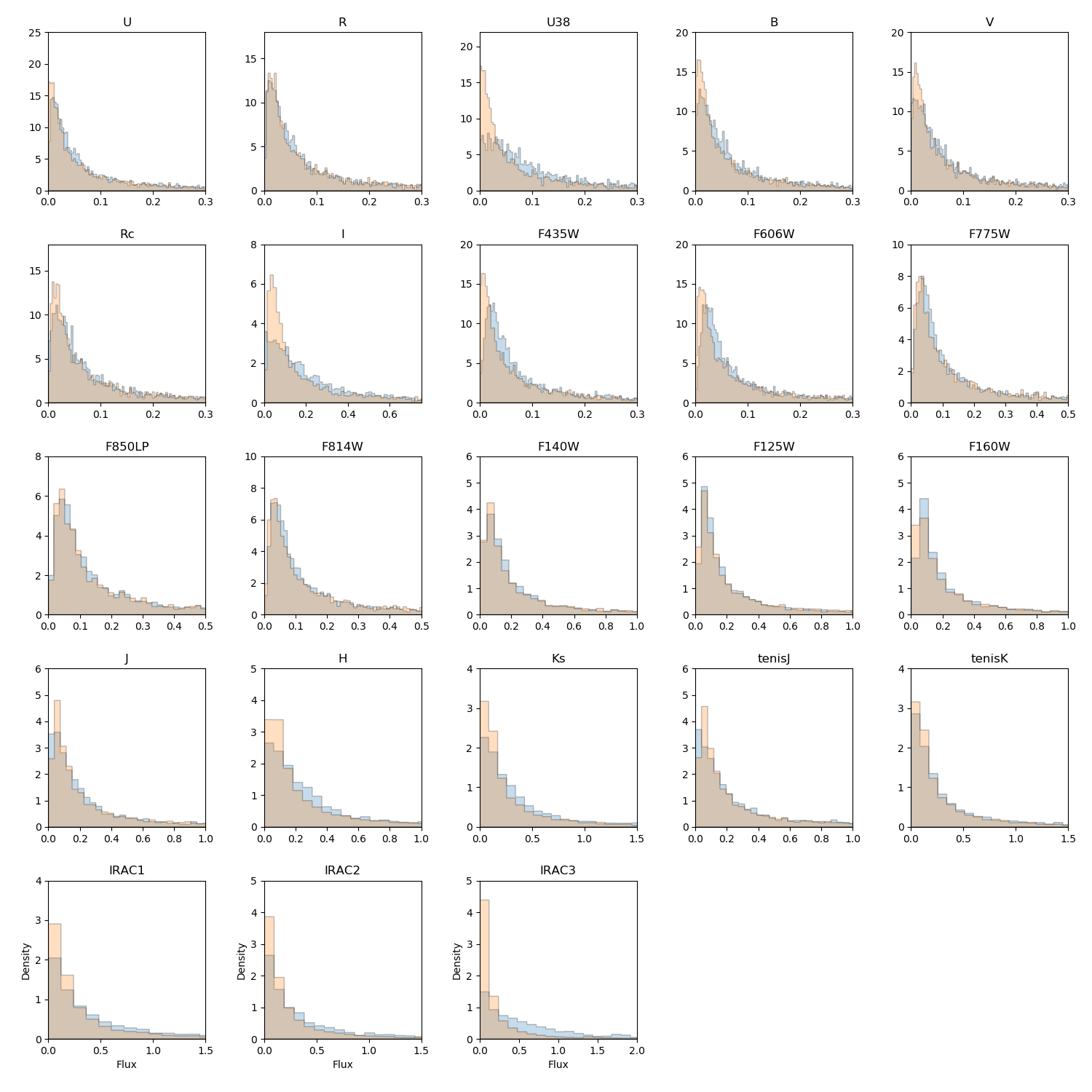}
    \caption{The blue histogram represents the distribution of observed flux values for each band in the reference 3D-HST sample described in the text. The orange histogram represents the distribution of flux values for each band in our simulated catalog, obtained using \protect\textsc{bagpipes} to generate the photometry.}
    \label{fig:Flux_Mock_Data_KDE_Params_Panel_5_5_z1}
\end{figure*}

We next compare the distribution of observational fluxes in several bands generated from \textsc{bagpipes} using an exponentially declining ($\tau$) SFH model with input parameters log$_{10}(\text{mass})$, age, $\tau$, $A_v$, and $z$ sampled from the multivariate KDE in Sec. \ref{sec:KDE} with that of the observational flux values in several bands of the 3D-HST catalog from redshift range $0.9 < z < 1.1$. The $\tau$ SFH model in \textsc{bagpipes} was chosen because it was also the model used to fit the observations by FAST, and since the multivariate KDE of the parameter space was based on these FAST values, comparison of these distributions acts as an important validation step. The flux values in the 3D-HST catalog are scaled to an AB magnitude zeropoint of 25.0, so a scaling factor of $\frac{1}{10^{0.44}} = 0.3631$ was employed before comparing it to the \textsc{bagpipes} generated flux values, which are in microjanskys ($\mu Jy$). Fig. \ref{fig:Flux_Mock_Data_KDE_Params_Panel_5_5_z1} shows that the \textsc{bagpipes} forward model is able to generate generally realistic 3D-HST observational flux values at the population level. Table \ref{tab:mock_vs_3dhst} provides summary statistics on the flux distribution in each band; some differences can be attributed to the different IMF used by the two approaches (Kroupa and Chabrier, respectively).

\begin{table}
\footnotesize
\centering
  	\vspace{-3mm}
  \caption{Median values and representative percentiles, in each band, for the reference 3D-HST sample at z $\sim$ 1 described in the text and the simulated galaxies generated by \protect\textsc{bagpipes} using the KDE distributions from Sec. \protect\ref{sec:KDE}. We find good agreement between the observed and simulated catalog in all bands. Some residual differences can be attributed to the different IMF used by the two approaches (Kroupa and Chabrier, respectively).}
  \label{tab:mock_vs_3dhst}
\begin{tabular}{l c c c}
  \hline
  Band & Median & 16\% & 84\%\\
  \hline
U$_{\text{3D}-HST}$ & 0.0488 & 0.0127 & 0.2200\\
U$_{\text{BP}}$ & 0.0434 & 0.0097 & 0.2070\\
\hline
R$_{\text{3D}-HST}$ & 0.0572 & 0.0155 & 0.2929\\
R$_{\text{BP}}$ & 0.0557 & 0.0147 & 0.3005\\
\hline
U38$_{\text{3D}-HST}$ & 0.0842 & 0.0230 & 0.2872\\
U38$_{\text{BP}}$ & 0.0438 & 0.0097 & 0.2081\\
\hline
B$_{\text{3D}-HST}$ & 0.0539 & 0.0140 & 0.2320\\
B$_{\text{BP}}$ & 0.0440 & 0.0100 & 0.2163\\
\hline
V$_{\text{3D}-HST}$ & 0.0571 & 0.0149 & 0.2569\\
V$_{\text{BP}}$ & 0.0463 & 0.0111 & 0.2330\\
\hline
Rc$_{\text{3D}-HST}$ & 0.0643 & 0.0174 & 0.3104\\
Rc$_{\text{BP}}$ & 0.0540 & 0.0144 & 0.2879\\
\hline
I$_{\text{3D}-HST}$ & 0.1894 & 0.0494 & 0.8176\\
I$_{\text{BP}}$ & 0.1132 & 0.0349 & 0.6181\\
\hline
F435W$_{\text{3D}-HST}$ & 0.0552 & 0.0194 & 0.2222\\
F435W$_{\text{BP}}$ & 0.0445 & 0.0100 & 0.2156\\
\hline
F775W$_{\text{3D}-HST}$ & 0.0582 & 0.0207 & 0.2589\\
F775W$_{\text{BP}}$ & 0.0489 & 0.0126 & 0.2546\\
\hline
F606W$_{\text{3D}-HST}$ & 0.0920 & 0.0325 & 0.4542\\
F606W$_{\text{BP}}$ & 0.0914 & 0.0260 & 0.4962\\
\hline
F814W$_{\text{3D}-HST}$ & 0.1231 & 0.0411 & 0.6331\\
F814W$_{\text{BP}}$ & 0.1189 & 0.0369 & 0.6470\\
\hline
F850LP$_{\text{3D}-HST}$ & 0.1031 & 0.0354 & 0.5044\\
F850LP$_{\text{BP}}$ & 0.0993 & 0.0293 & 0.5400\\
\hline
F140W$_{\text{3D}-HST}$ & 0.1668 & 0.0551 & 0.8863\\
F140W$_{\text{BP}}$ & 0.1580 & 0.0491 & 0.9222\\
\hline
F125W$_{\text{3D}-HST}$ & 0.1606 & 0.0553 & 0.8934\\
F125W$_{\text{BP}}$ & 0.1527 & 0.0472 & 0.8806\\
\hline
F160W$_{\text{3D}-HST}$ & 0.1701 & 0.0622 & 0.9500\\
F160W$_{\text{BP}}$ & 0.1512 & 0.0485 & 0.8965\\
\hline
J$_{\text{3D}-HST}$ & 0.1757 & 0.0456 & 0.9135\\
J$_{\text{BP}}$ & 0.1523 & 0.0465 & 0.8834\\
\hline
H$_{\text{3D}-HST}$ & 0.2362 & 0.0609 & 1.0381\\
H$_{\text{BP}}$ & 0.1554 & 0.0499 & 0.9346\\
\hline
Ks$_{\text{3D}-HST}$ & 0.2822 & 0.0715 & 1.4908\\
Ks$_{\text{BP}}$ & 0.1740 & 0.0562 & 1.0789\\
\hline
tenisJ$_{\text{3D}-HST}$ & 0.1907 & 0.0439 & 1.0315\\
tenisJ$_{\text{BP}}$ & 0.1539 & 0.0473 & 0.9146\\
\hline
tenisK$_{\text{3D}-HST}$ & 0.2265 & 0.0537 & 1.4288\\
tenisK$_{\text{BP}}$ & 0.1739 & 0.0562 & 1.0794\\
\hline
IRAC1$_{\text{3D}-HST}$ & 0.3669 & 0.0708 & 2.1866\\
IRAC1$_{\text{BP}}$ & 0.1886 & 0.0590 & 1.1943\\
\hline
IRAC2$_{\text{3D}-HST}$ & 0.2908 & 0.0518 & 1.7301\\
IRAC2$_{\text{BP}}$ & 0.1426 & 0.0441 & 0.9038\\
\hline
IRAC3$_{\text{3D}-HST}$ & 0.5486 & 0.1055 & 2.5733\\
IRAC3$_{\text{BP}}$ & 0.1003 & 0.0306 & 0.6496\\
\end{tabular}
\end{table}

\section{Bayesian Parameter Estimation and Model Evidence Calculation}
\label{sec:evidence}
\subsection{Bayesian Inference}
\label{sec:bayes_inf}

The goal of our analysis, which is carried out within the realm of Bayesian statistics, is to find the right balance between keeping the model simple and providing a good fit to the data, a concept often known as Occam's razor. Quantitatively, this translates into applying the model comparison formalism. The evidence (the denominator in Bayes theorem, sometimes also called marginal likelihood) is sensitive to not only the complexity of the parameter and prior space, but also the quality of the fit. Therefore, it can be used as a model comparison tool: models that have larger evidence are preferable, not because they necessarily fit the data better, but because they strike the right balance between complexity and accuracy.

The hypothesis we test in the paper is whether the evidence (or more accurately, the ratio of the evidence of two models, called the Bayes factor) can be successfully used to ``recognize”, in a purely data-driven manner, the correct model. To this end, we generate simulated data for a 5-parameter or 6-parameter model (the ``correct model" in the remainder of the paper), and then we ask: if we were to fit the data with the correct model (5 or 6 parameters, respectively) or a more simplistic model (4 or 5 parameters), would the Bayes factor favor the more complex model? Would we be able to recognize the ``true” complexity using the Bayes factor? If our tests are successful, this indicates a possible path to derive the ``true” model complexity from the data, giving us a way to discriminate between useful, meaningful parameters, and ``noisy” ones that add complexity without sufficiently improving the quality of the fit.

We start with estimation of the posterior distribution given a model $\bold{M}$, its parameters $\boldsymbol{\Theta}$, and a set of data $\bold{D}$. We follow the notational convention of \citealt{Speagle2019}. We use Bayes Rule,

\begin{equation} \label{eq:3.1}
P(\boldsymbol{\Theta} | \bold{D}, \bold{M}) = \frac{P(\bold{D} | \boldsymbol{\Theta}, \bold{M}) P(\boldsymbol{\Theta} | \bold{M})}{P(\bold{D} | \bold{M})}
\end{equation}

\noindent where $P(\bold{D} | \boldsymbol{\Theta}, \bold{M})$ is the likelihood of the data given the parameters in the model, $P(\boldsymbol{\Theta} | \bold{M})$ is the prior for the parameters, and

\begin{equation} \label{eq:3.2}
P(\bold{D} | \bold{M}) = \int_{\boldsymbol{\Theta}_{\Omega}} P(\bold{D} | \boldsymbol{\Theta}, \bold{M}) P(\boldsymbol{\Theta} | \bold{M}) \ d \boldsymbol{\Theta}
\end{equation}

\noindent is the marginal likelihood, also known as the $\textit{evidence}$, integrated over the entire parameter space $\Omega$. Rewriting Bayes Rule using shorthand notation, we have
\begin{equation} \label{eq:3.3}
P(\boldsymbol{\Theta}_\bold{M}) = \frac{L(\boldsymbol{\Theta}_\bold{M})\pi(\boldsymbol{\Theta}_\bold{M})}{Z_\bold{M}}
\end{equation}

\noindent for a given model $\bold{M}$, where

\begin{equation} \label{eq:3.4}
Z_\bold{M} = \int_{\boldsymbol{\Theta}_{\Omega}} L(\boldsymbol{\Theta}_\bold{M})\pi(\boldsymbol{\Theta}_\bold{M}) \ d \boldsymbol{\Theta}
\end{equation}

The evidence $Z_\bold{M}$ can therefore be understood simply as the likelihood space multiplied by the prior space integrated over the entire parameter space.

\subsection{Nested Sampling}
\label{sec:ns}

Numerous Markov Chain Monte Carlo (MCMC) methods have been been used to successfully generate samples and their associated weights for posterior estimation. However, when the posterior is multimodal or there are significant parameter degeneracies, traditional MCMC methods often struggle due to algorithmic limitations, computational expense, or both. Moreover, calculating evidence to a sufficient degree of accuracy using MCMC methods is often challenging if not impossible (\eg \citealt{Feroz2008}).

Nested sampling is an alternative to MCMC sampling developed by \citealt{Skilling2006} that can perform both accurate evidence calculation and effective posterior estimation while avoiding the issues noted above. It harnesses the relationship between the likelihood and prior, and samples from the prior space subject to ever-increasing lower bounds on the likelihood. Each successful iteration of the nested sampling algorithm results in an estimate of the remaining prior ``volume" as well as a likelihood evaluation, which are used to calculate a fraction of the total evidence, and by extension a posterior sample. This process creates nested ``shells" of iso-likelihood contours, hence the name.

To start, we can rewrite \eqref{eq:3.4} as 

\begin{equation} \label{eq:3.12}
Z = \int_{\boldsymbol{\Theta}_{\Omega}} P(\bold{D} | \boldsymbol{\Theta}) P(\boldsymbol{\Theta}) \ d \boldsymbol{\Theta} = \int_{\boldsymbol{\Theta}_{\Omega}} L(\boldsymbol{\Theta}) \pi (\boldsymbol{\Theta}) \ d \boldsymbol{\Theta}.
\end{equation}

The estimate of \eqref{eq:3.12} is simply its expected value. Note that in mathematical statistics, for a positive random variable, the area above a cumulative distribution function (CDF) and below 1 is the expected value. In other words, if the probability density function (PDF) of $X$ is f and the CDF of $X$ is F, then 

\begin{equation} \label{eq:3.13}
E(X) \equiv \int_0^{\infty} (1 - F(x))dx = \int_0^{\infty} x f(x) dx
\end{equation}

Furthermore, let the CDF of $L(\boldsymbol{\Theta}) = \lambda$ above be defined as

\begin{equation} \label{eq:3.14}
F(\lambda) \equiv \int_{L(\boldsymbol{\Theta}) < \lambda} \pi(\boldsymbol{\Theta}) d \boldsymbol{\Theta}
\end{equation}

\noindent so estimating \eqref{eq:3.12} is equivalent to estimating $\int_0^{\infty} (1 - F(x))dx$, where $F$ is defined in \eqref{eq:3.14}.

Then if we define the prior volume $X$ as $dX = \pi(\boldsymbol{\Theta}) d \boldsymbol{\Theta}$, such that

\begin{equation} \label{eq:3.15}
X(\lambda) = 1 - F(\lambda) = \int_{L(\boldsymbol{\Theta}) > \lambda} \pi(\boldsymbol{\Theta}) d \boldsymbol{\Theta}
\end{equation}

\noindent (noting the change in equality in the integration bounds) then \eqref{eq:3.12} is equal to $\int_0^{\infty} X(\lambda) d \lambda$. Lastly, inverting $X(\lambda)$ and evaluating the integral of the inverse from 0 to 1, we have

\begin{equation} \label{eq:3.16}
\int_0^1 X^{-1}(a) = \int_{\boldsymbol{\Theta}_{\Omega}} L(\boldsymbol{\Theta}) \pi (\boldsymbol{\Theta}) \ d \boldsymbol{\Theta} = \int_0^1 L(X) dx.
\end{equation}

\noindent where $L(X)$, the inverse of \eqref{eq:3.15}, is a monotonically decreasing function of $X$. Thus the evidence integral can be written as 

\begin{equation} \label{eq:3.17}
Z = \int_0^1 L(X) dx
\end{equation}

\noindent and its calculation is recast from a difficult multi-dimensional integration over $\boldsymbol{\Theta}$ to an easier one-dimensional integration over $X$. 

The next step in evidence calculation involves computing the likelihoods $L_i = L(X_i)$ where $X_i$ decreases as $i = 0, 1, 2, \ldots, M$ so that

\begin{equation} \label{eq:3.18}
0 < X_M < \ldots < X_2 < X_1 < X_0 = 1.
\end{equation}

\noindent The evidence is then found using quadrature as

\begin{equation} \label{eq:3.19}
Z = \sum_{i=1}^M L(\boldsymbol{\Theta}_i) w_i
\end{equation}

\noindent where $w_i = \frac{1}{2}(X_{i - 1} - X_{i + 1})$ (\citealt{Feroz2009}).

\subsection{Calculating the Evidence}
\label{sec:calculate_evidence}

The process of calculating the evidence involves several steps. First, $N$ ``live" points are sampled from the prior space and corresponding likelihoods are calculated for each live point. The likelihoods are then ordered $L_0, L_1, \ldots, L_N$, where $L_0$ is the smallest likelihood value. We will assume henceforth that the prior volume at this point in the sampling is $X_0 = \bold{1}$ (i.e. the unit hypercube). Next, $L_0$ is removed from the set of live points and becomes a ``dead" point. A new point is then drawn from the prior space, and if its likelihood value is greater than new smallest likelihood value (\ie $L_{\text{new}} > L_{0_{\text{new}}}$), it is added to the pool of live points. If, on the other hand, the new point's likelihood value is less than the new smallest likelihood value (\ie $L_{\text{new}} < L_{0_{\text{new}}}$, then it is discarded and added to the pool of ``dead" points. At each new live point (each ``successful" iteration), a new iso-likelihood contour is defined with which to constrain future live points, the prior space $X$ shrinks, and the evidence value increases. 

Nested sampling is therefore a form of rejection sampling, and indeed the biggest challenge in nested sampling is sampling a new point from the prior space to be added to the pool of live points subject to the constraint $L_{\text{new}} > L_{0_{\text{new}}}$. The trick is to find a way to both efficiently sample from the prior volume defined in Equation \eqref{eq:3.5} and also to fully explore the likelihood space while satisfying this constraint.

\subsection{Posterior Estimation}
\label{sec:posteriors}

If we are interested in posterior samples, we use the dead points from the nested sampling process (the points with the lowest likelihood values). This is written as

\begin{equation} \label{eq:3.23}
p(\boldsymbol{\Theta}_i) = \frac{L(\boldsymbol{\Theta}_i)w_i}{Z}
\end{equation}

\noindent where $Z$ and $w_i$ are given in \ref{eq:3.19}. One can then construct relevant marginal distributions and descriptive statistics such as means and percentile values. 

\subsection{Parameter Estimation In SED Fitting Using MultiNest}
\label{sec:param_est_multinest}

We now return to our simulated data, where we used samples from the kernel density estimates of the  parameter space inferred by FAST for the 3D-HST catalog as inputs to our galaxy catalogs.

For the purposes of realistic Bayesian model fitting and parameter estimation, we include noise, modeled after the 3D-HST catalog, in the simulated observations created in Sec. \ref{sec:data}. This is achieved by sampling from the KDE of the observational error in each band and randomly adding or subtracting the sampled amount to/from the corresponding observational flux. The minimum KDE error sample values are restricted to each of the minimum observational error values of the 23 bands from the selection of 4,567 galaxies, and the maximum KDE error sample values are restricted to the 95th percentile of the KDE errors.

The original prior configuration is shown in Table \ref{tab:priors}. We note that age prior is adaptive, in the sense that the upper bound is restricted to be no larger than the age of the universe at the relevant redshift.

\begin{table*}
\begin{center}
\resizebox{\linewidth}{!}{
\begin{tabular}{|c|cccccc|}
\hline
& Parameter & \% of objects & \% of objects & median & scatter & OLF  \\
& & in 68\% region & in 95\% region & bias  & & \\ 
 \hline
0x & Mass$_{\text{burst}}$ & 63 & 88 & 0.01 & 0.06 & 0.0 \\
Noise Reduction & Age$_{\text{delayed} \ \tau}$ & 63/45 & 89/69 & 0.00/-0.01 & 0.05/0.06 & 0.00/0.00 \\
& Mass$_{\text{delayed} \ \tau}$ & 56/30 & 86/53 & -0.01/0.02 & 0.06/0.04 & 0.00/0.01 \\
& $\tau$ & 66/46 & 90/69 & 0.04/0.01 & 0.09/0.11 & 0.09/0.11 \\
& $A_v$ & 62/60 & 89/85 & 0.00/0.01 & 0.13/0.14 & 0.12/0.17 \\
 \hline
5x & Mass$_{\text{burst}}$ & 47 & 74 & 0.00 & 0.05 & 0.00 \\
Noise Reduction & Age$_{\text{delayed} \ \tau}$ & 56/20 & 83/36 & 0.00/-0.03 & 0.04/0.05 & 0.01/0.00 \\
& Mass$_{\text{delayed} \ \tau}$ & 50/12 & 77/24 & 0.00/0.02 & 0.04/0.02 & 0.00/0.00 \\
& $\tau$ & 50/17 & 81/36 & 0.02/-0.05 & 0.08/0.09 & 0.07/0.04 \\
& $A_v$ & 63/48 & 87/70 & 0.00/0.00 & 0.10/0.12 & 0.08/0.13 \\
\hline
10x & Mass$_{\text{burst}}$ & 44 & 73 & 0.00 & 0.04 & 0.00 \\
Noise Reduction & Age$_{\text{delayed} \ \tau}$ & 50/12 & 76/24 & 0.00/-0.03 & 0.03/0.05 & 0.01/0.00 \\
& Mass$_{\text{delayed} \ \tau}$ & 46/7 & 76/14 & 0.00/0.01 & 0.03/0.01 & 0.00/0.00 \\
& $\tau$ & 48/12 & 79/25 & 0.01/-0.06 & 0.07/0.08 & 0.05/0.01 \\
& $A_v$ & 62/42 & 86/59 & 0.00/0.00 & 0.08/0.11 & 0.06/0.14 \\
\hline
20x & Mass$_{\text{burst}}$ & 45 & 71 & 0.00 & 0.03 & 0.00 \\
Noise Reduction & Age$_{\text{delayed} \ \tau}$ & 49/9 & 74/15 & 0.00/-0.03 & 0.03/0.04 & 0.01/0.00 \\
& Mass$_{\text{delayed} \ \tau}$ & 46/3 & 74/8 & 0.00/0.01 & 0.02/0.01 & 0.00/0.00 \\
& $\tau$ & 45/9 & 76/18 & 0.00/-0.06 & 0.06/0.07 & 0.03/0.00 \\
& $A_v$ & 62/32 & 87/47 & 0.00/0.01 & 0.07/0.11 & 0.04/0.15 \\
\hline
\end{tabular}}
\caption{This table shows the parameter fits for Scenario 1, which includes fits and summary statistics for both the 5 parameter values (the first value in a given column) and 4 parameter model (the second value in a given column).}
\label{tab:param_fits_scenario_1}
\end{center}
\end{table*}

\begin{table*}
\begin{center}
\resizebox{\linewidth}{!}{
\begin{tabular}{|c|cccccc|}
\hline
& Parameter & \% of objects & \% of objects & median & scatter & OLF  \\
& & in 68\% region & in 95\% region & bias  & & \\ 
 \hline
 & Age$_{\text{burst}}$ & 66 & 95 & 0.02 & 0.03 & 0.02 \\
0x & Mass$_{\text{burst}}$ & 73/55 & 94/78 & 0.03/-0.01 & 0.06/0.07 & 0.00/0.00 \\
Noise Reduction & Age$_{\text{delayed} \ \tau}$ & 69/61 & 93/85 & -0.01/0.01 & 0.05/0.05 & 0.00/0.01 \\
& Mass$_{\text{delayed} \ \tau}$ & 69/56 & 92/79 & -0.02/0.00 & 0.05/0.06 & 0.00/0.00 \\
& $\tau$ & 65/62 & 90/84 & 0.05/0.03 & 0.09/0.09 & 0.12/0.08 \\
& $A_v$ & 64/61 & 87/86 & 0.00/0.01 & 0.14/0.15 & 0.13/0.17 \\
 \hline
 & Age$_{\text{burst}}$ & 56 & 88 & 0.01 & 0.03 & 0.00 \\
5x & Mass$_{\text{burst}}$ & 62/34 & 87/53 & 0.02/-0.05 & 0.05/0.08 & 0.00/0.00 \\
Noise Reduction & Age$_{\text{delayed} \ \tau}$ & 59/36 & 87/62 & 0.00/0.00 & 0.04/0.05 & 0.00/0.01 \\
& Mass$_{\text{delayed} \ \tau}$ & 62/35 & 86/55 & -0.01/0.01 & 0.04/0.03 & 0.00/0.00 \\
& $\tau$ & 60/39 & 87/64 & 0.03/0.00 & 0.08/0.09 & 0.06/0.06 \\
& $A_v$ & 56/54 & 84/78 & 0.00/0.00 & 0.11/0.12 & 0.08/0.13 \\
\hline
 & Age$_{\text{burst}}$ & 57 & 87 & 0.01 & 0.03 & 0.00 \\
10x & Mass$_{\text{burst}}$ & 55/22 & 84/37 & 0.01/-0.07 & 0.05/0.08 & 0.00/0.00 \\
Noise Reduction & Age$_{\text{delayed} \ \tau}$ & 55/25 & 85/50 & 0.00/0.00 & 0.04/0.04 & 0.00/0.01 \\
& Mass$_{\text{delayed} \ \tau}$ & 56/26 & 85/44 & 0.00/0.01 & 0.04/0.02 & 0.00/0.00 \\
& $\tau$ & 53/27 & 84/52 & 0.02/-0.01 & 0.07/0.08 & 0.05/0.05 \\
& $A_v$ & 56/50 & 84/74 & 0.00/0.00 & 0.10/0.11 & 0.08/0.11 \\
\hline
 & Age$_{\text{burst}}$ & 59 & 87 & 0.01 & 0.03 & 0.00 \\
20x & Mass$_{\text{burst}}$ & 50/14 & 82/24 & 0.00/-0.10 & 0.05/0.08 & 0.00/0.00 \\
Noise Reduction & Age$_{\text{delayed} \ \tau}$ & 53/21 & 81/37 & 0.00/0.00 & 0.04/0.04 & 0.01/0.01 \\
& Mass$_{\text{delayed} \ \tau}$ & 53/16 & 83/27 & 0.00/0.01 & 0.03/0.02 & 0.00/0.00 \\
& $\tau$ & 49/19 & 82/35 & 0.01/-0.01 & 0.07/0.07 & 0.04/0.03 \\
& $A_v$ & 55/45 & 80/67 & 0.00/0.00 & 0.09/0.10 & 0.06/0.10 \\
\hline
\end{tabular}}
\caption{This table shows the parameter fits for Scenario 2, which includes fits and summary statistics for both the 6 parameter values (the first value in a given column) and 5 parameter model (the second value in a given column).}
\label{tab:param_fits_scenario_2}
\end{center}
\end{table*}

\begin{table*}
\begin{center}
\resizebox{\linewidth}{!}{
\begin{tabular}{|c|cccccc|}
\hline
& Parameter & \% of objects & \% of objects & median & scatter & OLF  \\
& & in 68\% region & in 95\% region & bias  & & \\ 
 \hline
 & Age$_{\text{burst}}$ & 66 & 95 & 0.02 & 0.03 & 0.02 \\
0x & Mass$_{\text{burst}}$ & 73 & 94 & 0.03 & 0.06 & 0.00 \\
Noise Reduction & Age$_{\text{delayed} \ \tau}$ & 69/46 & 93/73 & -0.01/0.00 & 0.05/0.06 & 0.00/0.01 \\
& Mass$_{\text{delayed} \ \tau}$ & 69/56 & 92/79 & -0.02/0.02 & 0.05/0.04 & 0.00/0.01 \\
& $\tau$ & 65/49 & 90/74 & 0.05/0.03 & 0.09/0.10 & 0.12/0.11 \\
& $A_v$ & 64/63 & 87/87 & 0.00/0.01 & 0.14/0.14 & 0.13/0.16 \\
 \hline
 & Age$_{\text{burst}}$ & 56 & 88 & 0.01 & 0.03 & 0.00 \\
5x & Mass$_{\text{burst}}$ & 62 & 87 & 0.02 & 0.05 & 0.00 \\
Noise Reduction & Age$_{\text{delayed} \ \tau}$ & 59/28 & 87/49 & 0.00/0.00 & 0.04/0.05 & 0.00/0.00 \\
& Mass$_{\text{delayed} \ \tau}$ & 62/15 & 86/32 & -0.01/0.02 & 0.04/0.01 & 0.00/0.00 \\
& $\tau$ & 60/28 & 87/50 & 0.03/0.01 & 0.08/0.09 & 0.06/0.05 \\
& $A_v$ & 56/51 & 84/75 & 0.00/0.00 & 0.11/0.12 & 0.08/0.13 \\
\hline
 & Age$_{\text{burst}}$ & 57 & 87 & 0.01 & 0.03 & 0.00 \\
10x & Mass$_{\text{burst}}$ & 55 & 84 & 0.01 & 0.05 & 0.00 \\
Noise Reduction & Age$_{\text{delayed} \ \tau}$ & 55/19 & 85/39 & 0.00/0.00 & 0.04/0.04 & 0.00/0.00 \\
& Mass$_{\text{delayed} \ \tau}$ & 56/9 & 85/18 & 0.00/0.02 & 0.04/0.01 & 0.00/0.00 \\
& $\tau$ & 53/21 & 84/41 & 0.02/-0.01 & 0.07/0.07 & 0.05/0.04 \\
& $A_v$ & 56/49 & 84/70 & 0.00/0.00 & 0.10/0.11 & 0.08/0.12 \\
\hline
 & Age$_{\text{burst}}$ & 59 & 87 & 0.01 & 0.03 & 0.00 \\
20x & Mass$_{\text{burst}}$ & 50 & 82 & 0.00 & 0.05 & 0.00 \\
Noise Reduction & Age$_{\text{delayed} \ \tau}$ & 53/16 & 81/30 & 0.00/0.00 & 0.04/0.03 & 0.01/0.00 \\
& Mass$_{\text{delayed} \ \tau}$ & 53/6 & 83/11 & 0.00/0.02 & 0.03/0.01 & 0.00/0.00 \\
& $\tau$ & 49/15 & 82/28 & 0.01/-0.02 & 0.07/0.07 & 0.04/0.02 \\
& $A_v$ & 55/43 & 80/65 & 0.00/0.00 & 0.09/0.10 & 0.06/0.12 \\
\hline
\end{tabular}}
\caption{This table shows the parameter fits for Scenario 3, which includes fits and summary statistics for both the 6 parameter values (the first value in a given column) and 4 parameter model (the second value in a given column).}
\label{tab:param_fits_scenario_3}
\end{center}
\end{table*}

\begin{table*}
\begin{center}
\resizebox{\linewidth}{!}{
\begin{tabular}{|c|cccccc|}
\hline
& Parameter & \% of objects & \% of objects & median & scatter & OLF  \\
& & in 68\% region & in 95\% region & bias  & & \\ 
 \hline
0x & Mass$_{\text{burst}}$ & 63/54 & 88/78 & 0.01/0.01 & 0.06/0.06 & 0.00/0.01 \\
Noise Reduction & Age$_{\text{delayed} \ \tau}$ & 63/36 & 89/70 & 0.00/0.04 & 0.05/0.06 & 0.00/0.06 \\
& Mass$_{\text{delayed} \ \tau}$ & 56/56 & 86/79 & -0.01/0.00 & 0.06/0.06 & 0.00/0.01 \\
& $\tau$ & 66/7 & 90/20 & 0.04/0.16 & 0.09/0.10 & 0.09/0.55 \\
& $A_v$ & 62/28 & 89/48 & 0.00/0.05 & 0.13/0.14 & 0.12/0.37 \\
 \hline
5x & Mass$_{\text{burst}}$ & 47/27 & 74/53 & 0.00/0.01 & 0.05/0.04 & 0.00/0.00 \\
Noise Reduction & Age$_{\text{delayed} \ \tau}$ & 56/26 & 83/52 & 0.00/0.03 & 0.04/0.05 & 0.01/0.05 \\
& Mass$_{\text{delayed} \ \tau}$ & 50/37 & 77/64 & 0.00/-0.01 & 0.04/0.04 & 0.00/0.00 \\
& $\tau$ & 50/9 & 81/20 & 0.02/0.15 & 0.08/0.1 & 0.07/0.49 \\
& $A_v$ & 63/24 & 87/52 & 0.00/0.03 & 0.10/0.11 & 0.08/0.30 \\
\hline
10x & Mass$_{\text{burst}}$ & 44/22 & 73/38 & 0.00/0.01 & 0.04/0.04 & 0.00/0.00 \\
Noise Reduction & Age$_{\text{delayed} \ \tau}$ & 50/21 & 76/38 & 0.00/0.03 & 0.03/0.05 & 0.01/0.05 \\
& Mass$_{\text{delayed} \ \tau}$ & 46/32 & 76/56 & 0.00/0.00 & 0.03/0.04 & 0.00/0.00 \\
& $\tau$ & 48/11 & 79/21 & 0.01/0.14 & 0.07/0.11 & 0.05/0.44 \\
& $A_v$ & 62/24 & 86/51 & 0.00/0.02 & 0.08/0.10 & 0.06/0.25 \\
\hline
20x & Mass$_{\text{burst}}$ & 45/22 & 71/38 & 0.00/0.01 & 0.03/0.04 & 0.00/0.00 \\
Noise Reduction & Age$_{\text{delayed} \ \tau}$ & 49/21 & 74/39 & 0.00/0.03 & 0.03/0.05 & 0.01/0.05 \\
& Mass$_{\text{delayed} \ \tau}$ & 46/33 & 74/56 & 0.00/0.00 & 0.02/0.04 & 0.00/0.00 \\
& $\tau$ & 45/9 & 76/21 & 0.00/0.14 & 0.06/0.11 & 0.03/0.44 \\
& $A_v$ & 62/32 & 87/50 & 0.00/0.02 & 0.07/0.10 & 0.04/0.25 \\
\hline
\end{tabular}}
\caption{This table shows the parameter fits for Scenario 1 vs Scenario 4, which includes fits and summary statistics for both the 5 parameter values with log$_{10}$ priors (the first value in a given column) and 5 parameter model with linear priors (the second value in a given column).}
\label{tab:param_fits_scenario_1_vs_4}
\end{center}
\end{table*}

\begin{table*}
\begin{center}
\begin{tabular}{l c c c}
  \hline
  Parameter & Prior & Units & Sampling \\
  \hline
$A_v$ & (0.0001, 3.0)  & Magnitudes & Uniform in log$_{10}$\\
age & (0.05, age$_\text{univ}$) & Gyr & Uniform in log$_{10}$\\
burst age & (0.05, 0.90) & Gyr & Uniform\\
$\tau$ & (0.01, 10.0) & Gyr & Uniform in log$_{10}$\\
log$_{10}$, Mass & (6.5, 12.5) & $M/M_\odot$ & Uniform  \\
log$_{10}$, Burst mass & (6.5, 12.5) & $M/M_\odot$ & Uniform \\
\hline
\end{tabular}
\caption{Parameter priors used for SED fitting in Scenarios 1, 2 and 3. We note that the prior used for the mass variables is effectively a uniform prior in $\text{log}_{10} \ M/M_\odot$.}
\label{tab:priors}
\end{center}
\end{table*}

\section{Bayes Factors in Model Comparison}
\label{sec:bayesfactors}

\subsection{Calculating Evidence}
\label{sec:calculate_evidence}

We can use the evidence values of a given pair of models, along with prior beliefs about the models, to calculate the Bayes factor. We have

\begin{equation} \label{eq:3.5}
Z_\bold{M} \equiv P(\bold{D} | \bold{M}) = \int_{\boldsymbol{\Theta}_{\Omega}} P(\bold{D} | \boldsymbol{\Theta}, \bold{M}) P(\boldsymbol{\Theta} | \bold{M}) \ d \boldsymbol{\Theta}
\end{equation}

\noindent where $Z_{\bold{M}_i}$ is the evidence and $\pi(M_i)$ is the prior belief for $M_i$, $i=(0,1)$. Henceforth we will assume $\pi(M_0) = \pi(M_1) = \frac{1}{2}$, so the Bayes factor is 

\begin{equation} \label{eq:3.6}
B_{01} = \frac{P(\bold{D} | \bold{M}_0)}{P(\bold{D} | \bold{M}_1)} = \frac{Z_{\bold{M}_0}}{Z_{\bold{M}_1}}
\end{equation}

Interpretation of the Bayes factor can be done through the Jeffreys' scale, which is an empirically calibrated scale (adapted from \citealt{Trotta2017}, \citealt{Jeffreys1998}). Small absolute values of $B_{01}$ between model $\bold{M}_0$ and model $\bold{M}_1$ indicate that the data and prior structure do not prefer either model. Large negative values express a preference for the $\bold{M}_0$ model (in our formalism, the simpler model), large positive values denote a preference for the $\bold{M}_1$ model (in our formalism, the more complex model). The reference threshold values we use are listed in Table \ref{tab:jeffreys}.

\begin{table}
\begin{center}
\begin{tabular}{l l l l} 
  \hline
  $|\ln B_{01}|$ & Odds & Probability & Strength of evidence \\
  \hline
 $< 1.0$ & $< 3:1$ & $< 0.750$ & Inconclusive \\
 $1.0$ & $\sim 3:1$ & $0.750$ & Weak evidence \\
 $2.5$ & $\sim 12:1$ & $0.923$ & Moderate evidence \\
 $5.0$ & $\sim 150:1$ & $0.993$ & Strong evidence \\
  \hline
\end{tabular}
\caption{Bayes factors are used to compare the evidence of two competing models $\bold{M}_0$ and $\bold{M}_1$. The table includes rule-of-thumb threshold values known as Jeffreys' Scale, and the probability column is in reference to posterior probabilities of the model that is favored.}
\label{tab:jeffreys}
\end{center}
\end{table}

\subsection{Bayes Factors in Savage-Dickey Density Ratio}
\label{sec:bf_sddr}

The Savage-Dickey Density Ratio (SDDR) is an analytical approach to calculating Bayes factors for model selection (\citealt{Dickey1971}, \citealt{Verdinelli1995}) and was first introduced for use in a cosmological context by \citealt{Trotta2007}. Given an $N$ parameter model $M_1$ (where $N \geq 2$) with free parameters $\boldsymbol{\theta} = \theta_1, \ldots, \theta_N$ and a nested $N - 1$ parameter model $M_0$ with free parameters $\theta_1, \ldots, \theta_{N-1}$ and fixed $\theta_N = \theta_\star$, the SDDR can be calculated using \ref{eq:3.6} by

\begin{equation} \label{eq:3.7}
B_{01} = \left.\frac{P(\boldsymbol{\theta} | \ \textbf{D}, \bold{M}_1)}{\pi(\bold{\boldsymbol{\theta}} | \ \bold{M}_1)} \right|_{\theta_N = \theta_\star}
\end{equation}

Evaluation of the SDDR therefore requires estimation of the PDF for both the marginalized posterior and the prior of $\theta_N$ to calculate the height of the densities at $\theta_\star$. Logspline density estimation (\citealt{Stone1986}), in which the logarithm of a PDF is modeled using polynomial spline, has been found to perform well for estimation of the SDDR under general conditions (\citealt{Wagenmakers2010}) and tends to outperform kernel-based density estimation (\citealt{Wetzels2009}). We will use the R package \texttt{logspline} (\citealt{Stone1997}) to calculate the logspline density estimates in order to derive the SDDR.

The SDDR requires that the two competing models $M_0$ and $M_1$ are nested and that the priors are separable, i.e.

\begin{equation} \label{eq:3.8}
\pi(\boldsymbol{\theta} | M_1 ) = \pi(\theta_1| M_1 ) \ \pi(\theta_2| M_1 ) \ \ldots \ \pi(\theta_N | M_0 ).
\end{equation}

In addition, the prior for the fitted parameters of $M_1$ should equal the prior for the fitted parameters under $M_0$ in order for $M_1$ to reduce to $M_0$.  Importantly, in order to effectively calculate the SDDR, a sufficient number of posterior samples are needed in the neighborhood of $\theta_\star$. We emphasize that, depending on the dimensionality of the SED model being fit, the number of samples and coverage of the posterior space scales with the number of live points used in \textsc{MultiNest}. To obtain sufficient coverage of posterior samples, we choose 1,500 live points. The sampling efficiency in MultiNest is inversely proportional to the enlargement factor of the  volume of the ellipsoid region. A greater enlargement factor therefore explores more completely the likelihood space. As we are interested in not only accurate evidence values but also accurate posterior estimates, we use a sampling efficiency of 0.25 instead of the default 0.3 in \textsc{MultiNest}.

We note that the Bayes factor is effectively the height of the posterior density of $\theta_N$ divided by the height of the prior density of $\theta_N$ evaluated at $\theta_\star$, the choice of prior has a great impact upon the derivation of the SDDR. This caveat notwithstanding, the computational resources needed to calculate the SDDR are minimal, so it is an appealing alternative to a nested sampling approach to calculating Bayes factors. 

\section{Results}
\label{sec:results}

We analyze different simulations scenarios in order to assess, on one hand, the impact of incorrect assumptions on the derived physical parameters, and on the other, the reliability of the Bayes factor as a tool to empirically discern the model complexity from the data. While the latter is inherently new to this paper, the former is a necessary validation step for the SED fitting process.

Our simulations involve the following approaches. In all cases, we begin by assuming a noise profile matching the one of the 3D-HST galaxies described earlier in paper. Since we found in our preliminary studies that S/N was the main factor driving the effectiveness of the Bayes factor, we then reduce the noise (maintaining the same profile) of the mock galaxy catalog by factors of 5, 10, and 20. The input parameters for the simulations are the same throughout this process to ensure a fair comparison in the presence of modeling systematics.

\subsection{How Many Stellar Populations?} 

In choosing our set of simulations, we are motivated by the question ``Can we detect multiple major episodes of star formation?". In fact, many physically motivated models of star formation, including semi-analytic models and hydrodynamical simulations (\eg \citealt{somerville_physical_2015}, \citealt{genel_introducing_2014}, \citealt{Pillepich2018}), show that the stellar assembly of galaxies might be well described by a ``smooth" component, describing the process of star formation in isolation, and a stochastic component, driven by interactive events such as mergers and winds. We describe each scenario in detail below.

\begin{figure*}
\centering
	\includegraphics[width=0.9\textwidth]{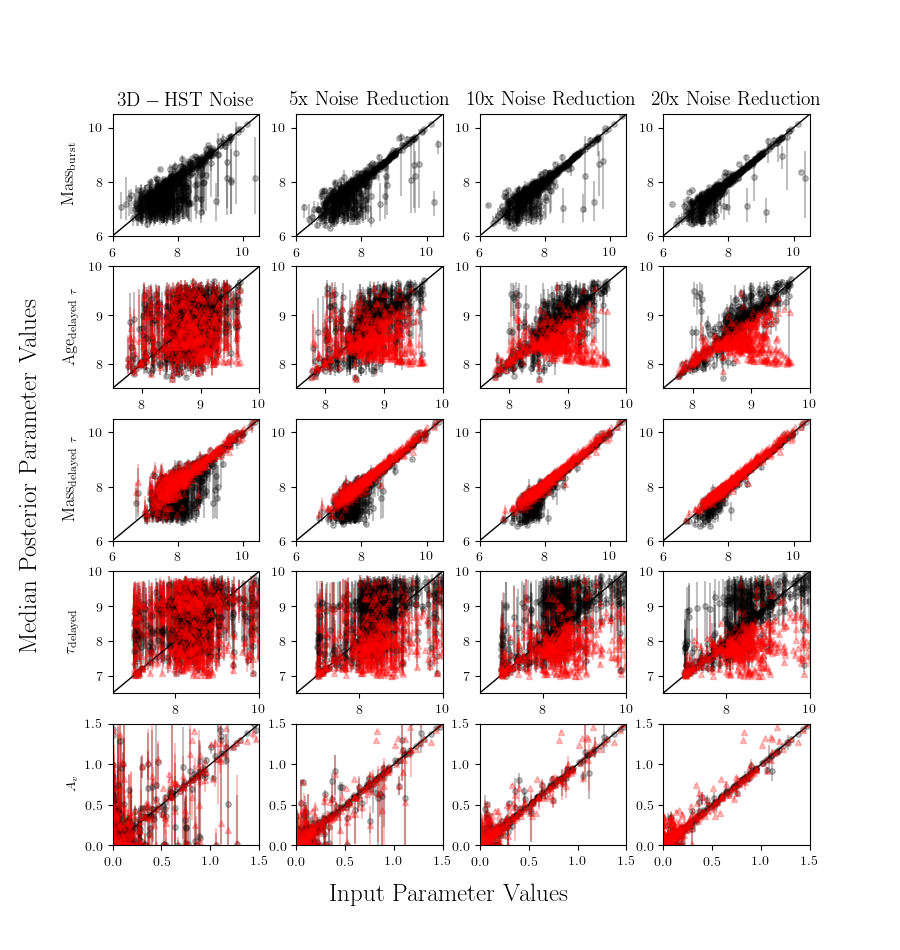}
    \caption{The black circles represent the input vs median posterior parameter values for the 5 parameter model in Scenario 1. The black vertical lines represent the 16\%-84\% credible intervals the parameters for the same model. The red triangles, on the other hand, are the input vs median posterior parameter values for the 4 parameter model in Scenario 1. The red vertical lines represent the 16\%-84\% credible intervals the parameters for the same model.}
    \label{fig:input_vs_median_params_new_scenario_1}
\end{figure*}

\begin{enumerate}
\item {\bf Scenario 1:} In our first test, we generate models using a five parameter model, with a ``main" stellar population component described by a delayed $\tau$ model, with parameters stellar age, stellar mass, $\tau$, and dust attenuation, and an additional recent burst component of star formation with age = 0.1 Gyr and variable stellar mass. We then perform parameter fits and evidence calculation both with a five parameter model ($\bold{M}_1$ for the purposes of Bayesian model comparison) with the parameters listed above, and a four parameter model where the burst mass is fixed at $\text{log}_{10} M/M_\odot$ = 6.5. In practice, the latter corresponds to a one-component stellar population, because the median mass for the entire sample is $\text{log}_{10} M/M_\odot \simeq 8.4$ and so the burst mass contribution to the total mass is at the percent level, but we can't simply set the mass in the burst to zero because that would not be part of the prior parameter space in the higher model, which is a necessary condition of nested models. These conditions aim to answer the question: ``Can we detect a recent burst of star formation?"

\item {\bf Scenario 2:} In our second test, we generate models using a six parameter model, where the ``main" stellar population component is still described by a delayed $\tau$ model, with parameters stellar age, stellar mass, $\tau$, and dust attenuation, and an additional burst component of star formation with variable age (between 0.05 and 0.9 Gyr) and variable stellar mass. We then perform parameter fits and evidence calculation both with a six parameter model ($\bold{M}_1$ for the purposes of Bayesian model comparison) with the parameters listed above, and a five parameter model where we vary the stellar mass, but fix the burst age at 0.1 Gyr. These conditions aim to answer the question: ``What happens if we incorrectly estimate the age of a secondary burst of star formation?"

\item {\bf Scenario 3:} Our third test involves generating models using the same six parameter model as in Scenario 2, and a four parameter model where the burst age is fixed at 0.1 Gyr and the burst mass is fixed at $\text{log}_{10} M/M_\odot$ = 6.75. These conditions aim to answer the question: ``Can we detect a secondary burst component of star formation, independent of age?"

\item {\bf Scenario 4:} Our fourth and last test attempts to test the impact of priors, in which we employ the same model as in Scenario 1 but with linear rather than $\text{log}_{10}$ priors on stellar age, $\tau$, and dust attenuation parameters.

\end{enumerate}

\subsection{Posterior Fits}

For scenarios 1-3, which involve different input parameter and fit parameters structure, we report the results of the parameter estimation procedure through a series of summary statistics for the parameter fits in Tables \ref{tab:param_fits_scenario_1}, \ref{tab:param_fits_scenario_2}, and \ref{tab:param_fits_scenario_3} following the conventions of \cite{Acquaviva2015}. For each parameter, we report the mean bias and scatter, as well as the fraction of outliers (defined as the percentage of objects for which the input and estimated values differ at more than 15\% level). Furthermore, in order to assess the accuracy of the reported posteriors, we report the fraction of objects for which the input values lie within the reported 68\%  and 95\% intervals.

For Scenario 1, we also show a set of scatter plots, displaying the input (true) values versus estimated values; results from the more complicated model are shown in black, and from the simpler model in red, in Fig. \ref{fig:input_vs_median_params_new_scenario_1}. The behavior of Scenarios 2 and 3 follow a very similar pattern.

Our results indicate that in all scenarios, the posterior width is underestimated when the simpler model is used. This becomes more stringent as the noise level is reduced: in the four parameter model (and at a modest level, even in the five parameter model) of Scenario 1, the size of the uncertainties, i.e. width of the posteriors, are very severely underestimated at the greatest amount of noise reduction.

In Scenarios 1-3, the bias, scatter, and outlier fraction for the simpler and more complicated models are comparable. They are generally present at the few percent level. Scenario 4 parameter fits are noteworthy for the significant outlier fraction for both $\tau$ and $A_v$ parameters, and significant bias in $\tau$ compared to those parameters Scenarios 1-3. Due to this, evidence values derived from Scenario 4 should not be considered reliable.

\subsection{Bayes Factors and Model Comparison}
\label{Sec:BF}

\begin{figure*}
\centering
	\includegraphics[width=0.8\textwidth]{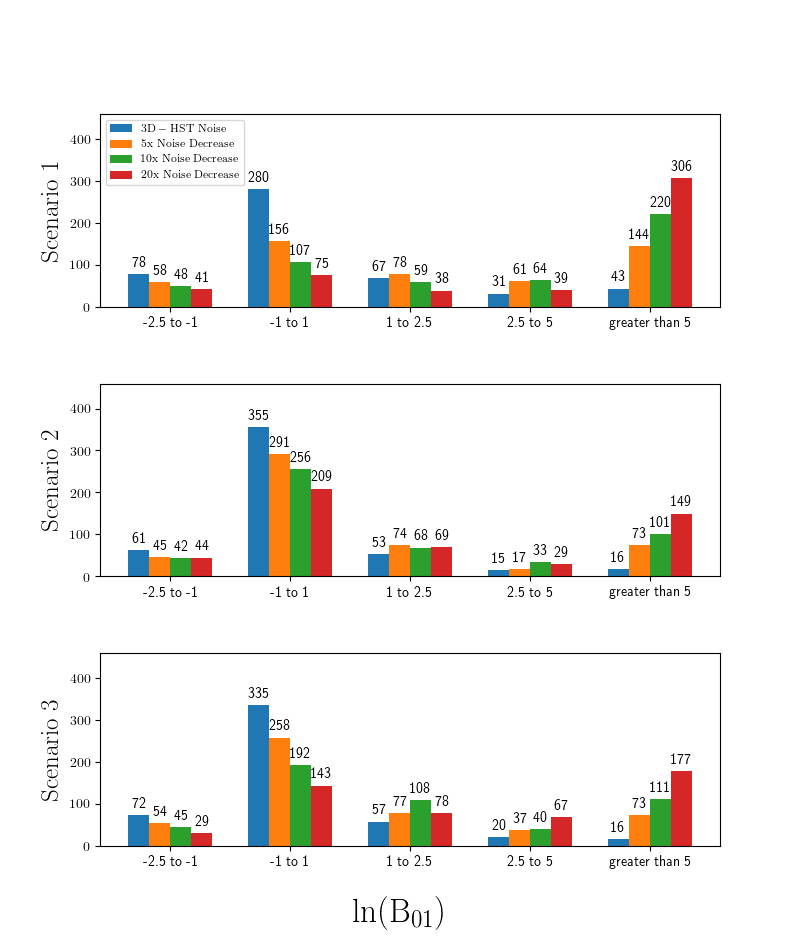}
    \caption{The top bar chart (Scenario 1) shows the Bayes Factor for data simulated using a 5-parameter model (delayed $\tau$ component + recent burst) and fitted with a 4-parameter model (delayed $\tau$ component), as described in the text. The middle bar chart (Scenario 2) shows the Bayes factor for data simulated using a 6-parameter model (delayed $\tau$ component + burst of varying mass/age) and fitted with a 5-parameter model (delayed $\tau$ component + recent burst). The bottom bar chart (Scenario 3) shows the Bayes factor for data simulated using a 6-parameter model (delayed $\tau$ component + burst of varying mass/age) and fitted with a 4-parameter model (delayed $\tau$ component). The horizontal axis show the distribution of the Bayes Factor for different S/N ratios, grouped according to the criteria of the Jeffrey Scale. From left to right, the ``columns" indicate that the BF moderately prefers the simpler model, is indecisive, moderately prefers the more complex model, decisively prefers the more complex model. In all scenarios, the Bayes Factor ``fails" (prefers the simpler model) in a minority of cases, especially at high S/N.}
    \label{fig:bf_jeffreys_1-2-3_new_scenarios}
\end{figure*}

We can now turn to the analysis of the Bayes factors, \ie the natural log ratio of evidences of the more complex (correct) vs simpler (incorrect) model for the three scenarios; our results are shown in  Fig. \ref{fig:bf_jeffreys_1-2-3_new_scenarios}. Each set of multi-histograms (``column") shows the Bayes factor; the columns are organized according to the Jeffrey scale described in Table \ref{tab:jeffreys}. The leftmost column displays cases in which the simpler model is preferred with moderate confidence ($-2.5 < \text{ln} \  B_{01} < -1$), and moving to the right, we find inconclusive cases, and cases in which the more complex model is preferred with weak, moderate, and high confidence respectively.

In an ideal case, the Bayes factor would always prefer the more complex model, since we know that this is the correct underlying model; however, the most crucial issue is whether the Bayes factor can actually be misleading, preferring the simpler (incorrect) model with high confidence.

In Scenario 1, at the 3D-HST noise level, we see that the Bayes factor is indecisive in the majority of cases (56\% of the total), and is able to ``pick" the more complex model with high confidence (ln $B_{01} > 2.5$) only in 15\% of the cases. Importantly, though, there are only a handful of failing cases ($\sim 6\%$ of the total, not shown) in which the Bayes factor strongly prefers the simpler model, and a fraction of cases (16\% of the total) in which the Bayes factors indicates a moderate preference for the simpler model. As the noise level is reduced, however, the predictive power of the evidence increases, and the Bayes factor indicates a strong preference for the more complex model in the majority of cases (57\% for a noise reduction of 10, and 69\% for a noise reduction of 20). We note that the latter constraints could also easily be recast as a condition on stellar mass of galaxies within the 3D-HST survey at $z \sim$ 1; roughly speaking, the Bayes factor favors the more complex model with high confidence for galaxies with a stellar mass  $\>= 2.5 \times 10^9$ $M_\odot$.

In Scenario 2, we observe that the Bayes factor remains a fair estimator of the preferred model, with no cases where the Bayes factor strongly favors for the simpler model, and a small fraction ($12\%$ of the total) of failures, in which the simpler model is weakly favored. Similar to scenario 1, at the 3D-HST noise level the Bayes factor is largely indecisive, but becomes more and more able to select the 6 parameter model with confidence as the noise is reduced, with a strong preference for the more complex model for $27\%$ and $36\%$ of galaxies in the case of a 10x and 20x noise reduction, respectively.

In Scenario 3, the failures tend to disappear more quickly as S/N increases, but otherwise we observe a similar trend as in Scenario 2 in the Bayes factor's ability to select the more complex 6 parameter model as the noise is reduced. At the 3D-HST noise level, the Bayes factor fails only $14\%$ of the time, in which the simpler model is weakly favored.  At the 10x and 20x noise reduction, the Bayes factor strongly prefers the more complex model $22\%$ and $35\%$, respectively.

Our results indicate that even in the presence of modeling systematics, the Bayes factor is a fairly reliable indicator of model complexity, and is able to detect a secondary episode of star formation in the vast majority of cases, especially at high S/N.

\subsection{Investigating Failures}

We now attempt to investigate of the ``failures" of the Bayes factor (i.e. when the Bayes factor incorrectly prefers the simpler model, or ln $B_{01} < -1$) as a tool to discern model complexity. We would like to understand, if possible, what factors may contribute to skew the evidence toward the simpler (incorrect) model, resulting in the simpler model being strongly or moderately preferred. We consider two possible explanations: the S/N ratio of each individual spectrum, which we expect to be important given the improvement seen at the population level with increasing S/N, and the age separation between the main and the secondary population. The rationale for the latter is that if the two population have similar ages, they effectively act as a single population, and in this sense the single-population model is still - in practical if not formal terms - the correct choice.

\begin{figure*}
\centering
	\includegraphics[width=0.8\textwidth]{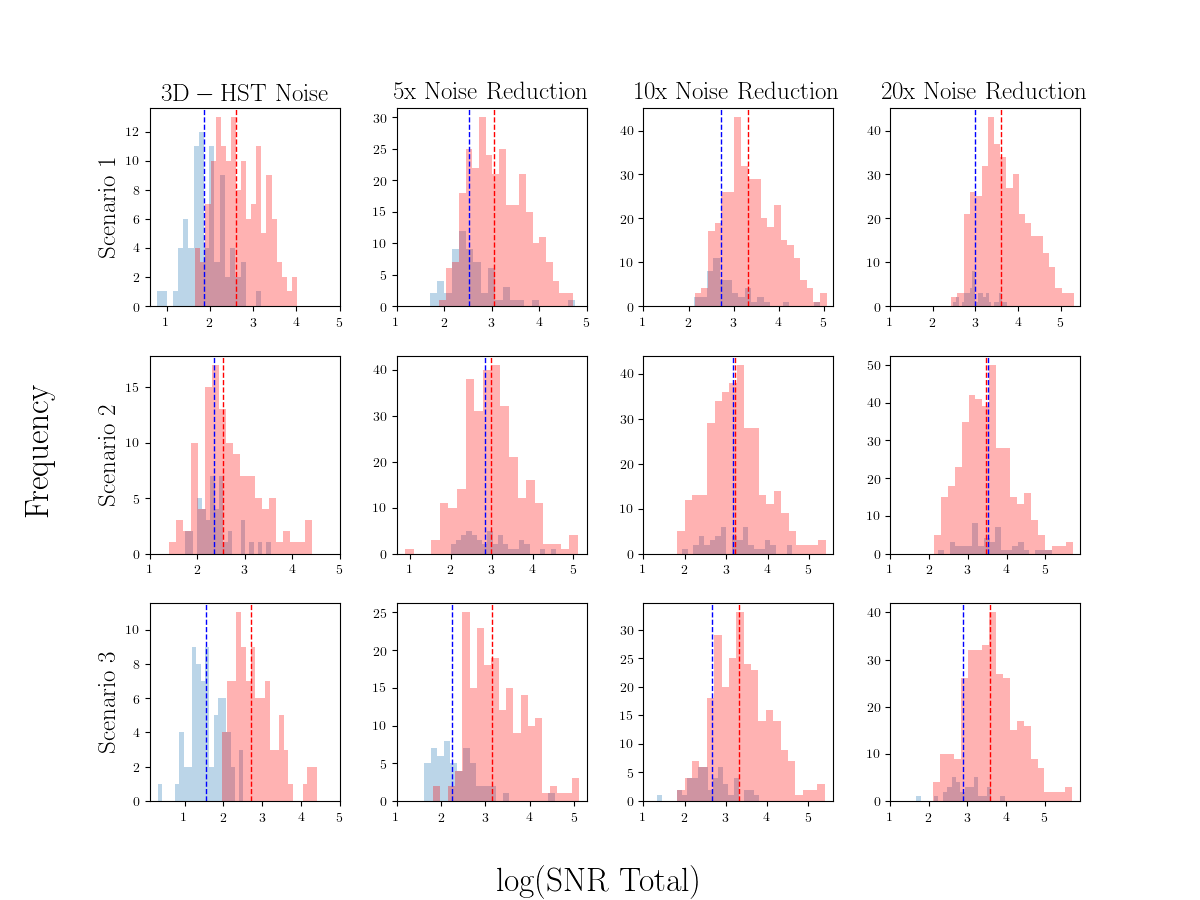}
    \caption{The blue histograms represent the distribution of cases in which the Bayes factor incorrectly prefers the simpler model (ln $B_{01} < -1$) and the red histograms represent the distribution of cases in which the Bayes factor correctly prefers the more complex model (ln $B_{01} > 1$). We show results for Scenarios 1-3 at each noise level, as a function of the total S/N of each SED. The total S/N is the sum of every band's photometric flux measurement divided by its 1$\sigma$ error measurement. The dotted vertical line represents the median value for each distribution. The S/N of each SED is an important factor in determining the method's success in Scenarios 1 and 3.}
    \label{fig:failing_vs_succeeding_bf_scenarios_1_2_3}
\end{figure*}

In Fig. \ref{fig:failing_vs_succeeding_bf_scenarios_1_2_3} we plot the distribution of cases in which the Bayes factor incorrectly prefers the simpler model (ln $B_{01} < -1$) and those is in which the Bayes factor correctly prefers the more complex model (ln $B_{01} > 1$), for each scenarios and noise level, as a function of the total S/N of each SED.

Scenarios 1 and 3 both show a significantly higher median total S/N for the succeeding cases in comparison to the failing cases, indicating that S/N is probably the main driver of the expected performance of the method. This is encouraging because Scenarios 1 and 3 correspond to a practical situation in which we are trying to detect a secondary population (recent, in Scenario 1, and of any age, for Scenario 2) by comparing a two-population fit with a single-population fit. From this plot, as well as the results presented in Sec. \ref{Sec:BF}, we can conclude that for sufficiently deep photometry, the Bayes factor is a reliable indicator of model complexity in the vast majority of cases.

The situation is not the same for Scenario 2, in which we attempt to fit two-population SEDs by fixing the age of the secondary population, and varying its mass. In this case, we see that the SED's total S/N is less correlated with the distribution of failing cases. This happens possibly because of increased degeneracies between the simpler (5-parameter) model and the more complex (6-parameter) model for this scenario, which make it harder to isolate one single discriminating factor. We note that this Scenario is the least akin to a physical situation in which we would want to use the Bayes Factor as a discriminating tool for model complexity.

\begin{figure*}
\centering
	\includegraphics[width=0.8\textwidth]{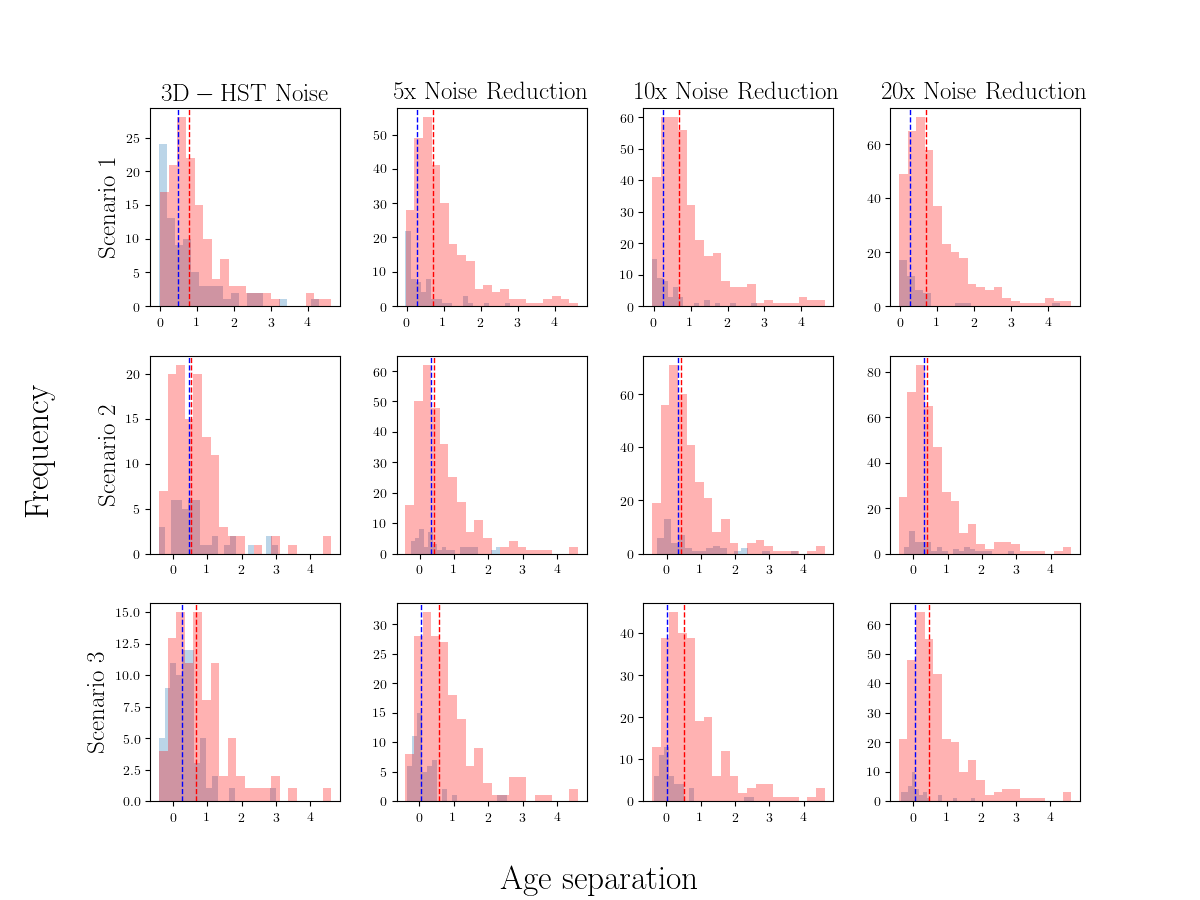}
    \caption{The blue histograms represent the distribution of cases in which the Bayes factor incorrectly prefers the simpler model (ln $B_{01} < -1$) and the red histograms represent the distribution of cases in which the Bayes factor correctly prefers the more complex model (ln $B_{01} > 1$), for each scenario and S/N level. The horizontal axis of each subplot is the age separation, which is defined as the time between the onset of star formation in the delayed $\tau$ SFH component and the burst SFH component (in Gyr). The dotted vertical line represents the median value for each distribution in each subplot.}
    \label{fig:failing_vs_succeeding_age_sep_scenarios_1_2_3}
\end{figure*}

A similar trend can be seen when we plot the same histograms of failing and succeeding cases as a function of age separation between the main (delayed $\tau$ SFH) component and the secondary (burst) component. Age separation here is defined as the time between the onset of star formation in the delayed $\tau$ SFH and the burst component. In Fig. \ref{fig:failing_vs_succeeding_age_sep_scenarios_1_2_3}, Scenarios 1 and 3 show a significant difference in the median age separation for the failing vs succeeding cases at each noise level; as expected, the Bayes factor is less able to pick the correct model if the age separation between the two components is small, because they effectively ``become one". Conversely, in Scenario 2 there is little difference in the median age separation for these two cases. 

\subsection{Design Space and Impact of Priors} 

\begin{figure*}
\centering
	\includegraphics[width=0.9\textwidth]{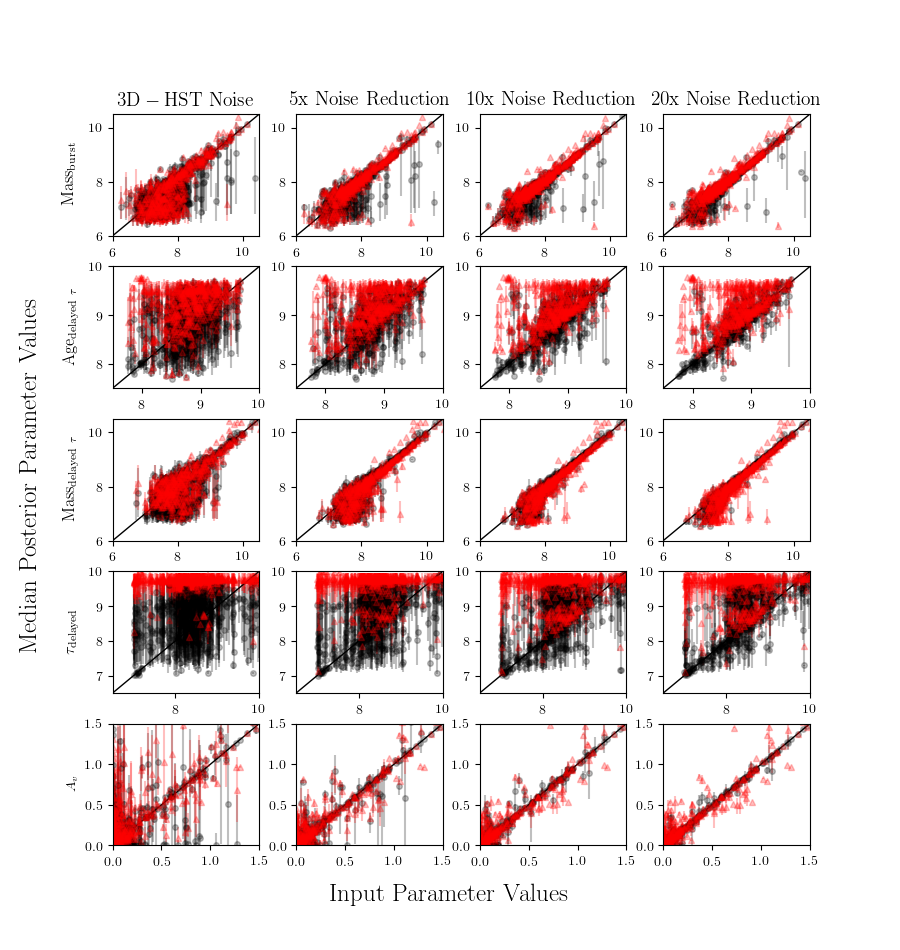}
    \caption{The black circles represent the input vs median posterior parameter values for the 5 parameter model in Scenario 1. The black vertical lines represent the 16\%-84\% credible intervals the parameters for the same model. The red triangles, on the other hand, are the input vs median posterior parameter values for the 5 parameter model in Scenario 4. The red vertical lines represent the 16\%-84\% credible intervals the parameters for the same model.}
    \label{fig:input_vs_median_params_new_scenario_1_vs_scenario_4}
\end{figure*}

One often neglected issue in Bayesian parameter estimation is the role that the choice of priors plays in determining the final results. We are interested in the recovery of physical parameters, as well as in how the choice of priors correlates with the reliability of the Bayes factor as a model selection tool.

We show a preliminary example (Scenario 4). We begin by comparing two five parameter models: one analog to the one described in Scenario 1, and one in which we employ a uniform prior in the age, $\tau$ parameter, and dust parameter $A_v$ of the main stellar population. We note that since the multivariate input parameter distributions are derived directly from the data (as opposed to being generated from a prior distribution), this test should be effective at determining which priors are preferable at a general level.

Fig. \ref{fig:input_vs_median_params_new_scenario_1_vs_scenario_4} and Table \ref{tab:param_fits_scenario_1_vs_4} show that the set of priors from Scenario 1 are highly preferable to the ones used here. The decay in the quality of the fits is evident, with the emergence of significant biases in some of the affected parameters, such as $A_v$ and $\tau$, which also present very large outlier fractions, of the order of 20-30 and 50\% respectively. Our recommendation from this simple test is that uniform priors in $\text{log}(A_v)$ and $\text{log}(\tau)$ should be used.

\begin{figure*}
\centering
	\includegraphics[width=0.8\textwidth]{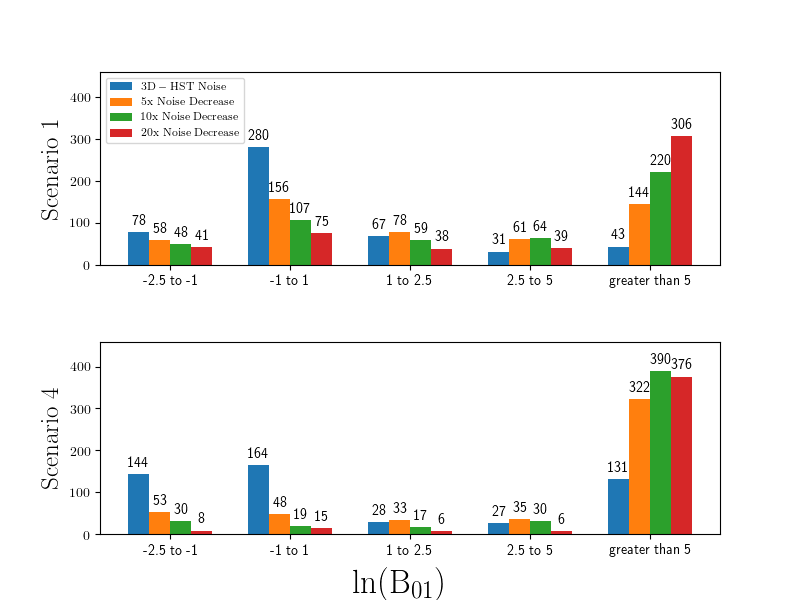}
    \caption{The top bar chart (Scenario 1) shows the  Bayes factor for data simulated using a 5-parameter model (delayed $\tau$ component + recent burst) and fitted with a 4-parameter model (delayed $\tau$ component). The bottom bar chart (Scenario 4) shows the Bayes factor for the same parameter space, but when uniform priors in $A_v$, Age$_{\text{delayed} \ \tau}$, and $\tau$ and are used, instead of uniform priors in the logarithm of the same variables, like in Scenario 1. Notation is the same as Fig. \ref{fig:bf_jeffreys_1-2-3_new_scenarios}. Despite the strong decay in the quality of parameter estimates observed in Scenario 4 (see Table \protect\ref{tab:param_fits_scenario_1_vs_4}), the Bayes factor remains reliable as a model complexity selection tool.}
    \label{fig:bf_jeffreys_1_vs_4_new_scenarios}
\end{figure*}

In Fig. \ref{fig:bf_jeffreys_1_vs_4_new_scenarios}, we show the Bayes factor from the comparison of a five parameter model, described in Scenario 1 above, and a five parameter model in which the priors are uniform in age, $\tau$, and $A_v$, as opposed as in log$_{10}$. 

The main result from this test is that at the 3D-HST noise level, priors play an important role in determining the Bayes factor, and the results obtained by using the uniform priors from Scenario 4 are not trustworthy in a higher number of cases. On the other hand, as the noise is reduced, the contribution of the data to the marginal likelihood increases, and the Bayes factor remains a reliable estimator of model complexity. \\

\section{Evidence vs. Savage-Dickey Density Ratio} 

Our last novel result comes from the comparison of the Bayes factor evaluated through the SDDR versus through the evidence. As mentioned above, the SDDR can be easily obtained by ordinary MCMC sampling, as opposed to the evidence calculation which requires expensive ad-hoc methods such as nested sampling. 

Table \ref{tab:sddr_table} provides a measure of the linear correlation between the natural log of the Bayes factor derived from the SDDR and the natural log of the Bayes factor derived from nested sampling in Scenarios 1-2 at each noise level. Scenario 3 is not included because it involves comparison of a 6 and 4 parameter model, and the SDDR method is valid only with $N$ vs $N-1$ model scenarios. Scenario 4 is also not included because the poor posterior fits in Sec. \ref{sec:posteriors} indicate that the evidence calculations are not reliable for model comparison. The first column ($\rho_{total}$) refers to the Pearson correlation coefficient calculated before any requirement is made on the number of samples near $\theta_{star}$, so it includes all 500 models for each scenario at each noise level. The second column ($\rho_{sub}$) lists the Pearson correlation coefficient calculated when considering on only models that have at least 1\% of their samples within $0.25 \times 10^9$ solar masses of $\theta_{\star}$. The third column lists the resulting number of models considered due to this requirement, and the fourth column lists the median number of posterior samples used to calculate $\rho_{sub}$.

These results show a high degree of correlation between the SDDR and nested sampling methods of Bayes factor computation, as long as the region of the posteriors where $\theta_{\star}$ is evaluated are sufficiently well sampled. They suggest that with dense sampling of the tails of the posterior and when the other conditions for the applicability of the SDDR approximation hold - most notably nested models - the SDDR can be successfully used as a proxy for the full calculation stemming from the evidence.

\begin{table}
\begin{center}
\begin{tabular}{c l l r r} 
  \hline
 & $\rho_{total}$ & $\rho_{sub}$ & models & median post\\
 &  &  & ($\rho_{sub}$) & samps ($\rho_{sub}$)\\
  \hline
 & 0.770 & 0.963 & 392 & 953 \\
Scenario 1 & 0.739 & 0.944 & 255 & 953 \\
Noise Levels & 0.779 & 0.952 & 186 & 1069 \\
& 0.757 & 0.932 & 139 & 1000 \\
  \hline
 & 0.761 & 0.928 & 472 & 1138 \\
Scenario 2 & 0.838 & 0.889* & 409 & 1109 \\
Noise Levels & 0.717 & 0.892* & 373 & 1267 \\
& 0.783 & 0.907 & 330 & 1321 \\
  \hline 
\end{tabular}
\caption{$\rho_{total}$ refers to the Pearson correlation coefficient derived from the Bayes factor calculated using either nested sampling or the SDDR, from the full set of 500 models. $\rho_{sub}$ refers to the Pearson correlation coefficient calculated using only a subset of models that have at least 1\% of their samples within $0.25 \times 10^9$ solar masses of $\theta_{\star}$. Numbers with * indicate that a single outlier was removed to calculate $\rho_{\text{sub}}$.}
\label{tab:sddr_table}
\end{center}
\end{table}

\section{Conclusions and future work}
\label{sec:conclusions}

In this paper we have explored several simulated scenarios in order to establish whether Bayesian model comparison can be relied upon as a method to estimate the ``true" model complexity afforded by the data. Our main conclusions are the following:
\begin{itemize}
    \item The Bayes factor is a promising tool to evaluate model complexity. In all the scenarios we have explored, the Bayes factor is able to select the correct model with high confidence in the  majority of cases, particularly at high S/N, while it ``fails", \ie selects the wrong model with high confidence, in only a few percent of the cases.  This is relevant in view of data from upcoming telescopes, for example the Nancy Grace Roman Space Telescope (NRST, formerly WFIRST) and the James Webb Space Telescope (JWST). When combined, data from these two observatories will provide deep coverage over the entire range of wavelengths (with the exception of the bluest channels, but extending to mid-infrared) considered here. While obtaining a direct comparison is difficult without making reference to a specific observing proposal, it can generally be expected that data from JWST and NRST will be 2-3 magnitudes deeper than those currently available from HST and Spitzer, for deep fields of area ~0.25 square degrees or larger (A. Koekemoer, in proceedings of NRST conference, Space Telescope, October 2020, and \citealt{koekemoer2019ultra}). A uniform improvement of 2.5 magnitude deeper data across the range of the spectrum would correspond to the ``10x noise reduction” case we have described in this paper.

    \item In the two scenarios that most closely resemble a comparison between a single component of star formation and a two-component model with an additional (recent or otherwise) burst, the Bayes factor performs increasingly well as a function of increasing S/N and age separation between the two components.
    \item Priors play a large role in parameter estimation and Bayes factors estimation, especially at low S/N, as is to be expected. Our recommendation is that uniform priors in the logarithms of stellar age, $A_v$, and $\tau$, are preferred to uniform priors in the same variables, although we also observe that the Bayes factor is less affected by incorrect priors than the estimates of physical parameters.
    \item The Savage Dickey Density Ratio can be a good proxy for the Bayes factor when the region of the posteriors where $\theta_{\star}$ is evaluated is sufficiently well sampled. This enables the usage of Bayesian model comparison even without the computationally intensive need for nested sampling algorithms, broadening the range of possible scientific applications of this method.
\end{itemize}
Finally, we believe that our Kernel Density Estimate method to extract realistic multivariate distributions for input parameters from data, keeping into account the covariances between different physical quantities, will be useful to the broader astrophysical community.

In a future paper, we aim to consider more carefully the sources of systematic uncertainty of the Bayes factor, and to run multiple simulations on each scenario to account for the stochasticity of the evidence calculation in the \textsc{MultiNest} nested sampling algorithm.

\section*{Acknowledgements}

We are grateful to Adam Carnall, Kartheik Iyer, and Eric Gawiser for many useful discussions, as well as to the anonymous referee for many useful comments that helped us improve the paper in content and form. This work was partially supported by a PSC-CUNY Type A award (62101-0050) and a Google Cloud Platform Research for Education grant. VA thanks the Institute of Cosmic Sciences of the University of Barcelona for hospitality during the completion of this work.





\bibliographystyle{mnras}
\bibliography{main.bib}








\bsp	
\label{lastpage}
\end{document}